\documentclass[useAMS,usenatbib]{mn2e}
\usepackage{times,psfig,epsfig,graphics,graphicx,multicol,amssymb} 

\def\aap{A\&A}
\def\gca{GeCoA}
\def\apj{ApJ}

\def\apjl{ApJ}
\def\mnras{MNRAS}

\def\nat{Nat}

\def\apss{Ap\&SS}

\def\lesssim{\mathrel{\hbox{\rlap{\hbox{\lower4pt\hbox{$\sim$}}}\hbox{$<$}}}}
\def\gesssim{\mathrel{\hbox{\rlap{\hbox{\lower4pt\hbox{$\sim$}}}\hbox{$>$}}}}

\begin{document}  
 \author[A. Morandi et al.]
{Andrea Morandi${}^1$\thanks{E-mail: andrea.morandi@studio.unibo.it}, Stefano Ettori${}^2$\\
$^{1}$ Dipartimento di Astronomia, Universit\`a di Bologna, via Ranzani 1,
I-40127 Bologna, Italy\\
$^{2}$ INAF-Osservatorio Astronomico di Bologna, via Ranzani 1, I-40127 Bologna, Italy}

\date{}

\title[Entropy profiles in X-ray luminous galaxy clusters at $z>0.1$]
{Entropy profiles in X-ray luminous galaxy clusters at $z>0.1$}
\maketitle

\begin{abstract}
The entropy distribution of the intracluster gas
reflects both accretion history of the gas and processes of feedback which provide a further
non-gravitational energy besides the potential one. 
In this work, we study the profiles and the scaling properties of the gas entropy
in 24 hot ($kT_{\rm gas} > 6$ keV) galaxy clusters observed with {\em Chandra}
in the redshift range 0.14--0.82 and showing different states of relaxation.
We recover the gas density, temperature and entropy profiles in a non-parametric way,
just relying on the assumption of a spherically symmetric emission in the deprojection
of the best-fit results of the spatially resolved X-ray spectral analysis.
Adding the hydrostatic equilibrium hypothesis, radial profiles are also obtained
from the deprojection of the surface brightness, allowing to verify whether 
the hydrostatic equilibrium is a tenable hypothesis by comparison with the spectral 
measurements. We confirm that this is the case on scales larger than $100$ kpc and
discuss the deviations observed in few non-cooling core clusters in the inner regions.
We show that the entropy profiles are remarkably similar outside the core and can be described by simple power-laws with slope of $1.0-1.2$. We measure an entropy level at $0.1 \, R_{200}$ of $100-500 \,\rm{keV \,cm^2}$ and a central plateau which spans a wide range of value ($\sim$ a few$-200\,\rm{keV \,cm^2}$) depending on the state of relaxation of the source.
The entropy values resolved at given fraction of the virial radius are proportional 
to the gas temperature in these hot systems and appear larger at higher redshift once 
they are compared to the local estimates.
To characterize the energetic of the central regions,
we compare the radial behaviour of the temperature of the gas with the temperature of the dark matter $T_{\rm DM}$ by estimating the excess of energy $\Delta E=3/2\, k(T_{\rm gas}- T_{\rm DM})$. We point out that $\Delta E$ ranges from $\approx 0$ in typical cooling-core clusters to few keV within $100$ kpc in non-cooling core systems. We also measure a significant correlation between the total iron mass and the entropy outside the cooling region,whereas in the inner regions they anti-correlate strongly. We find that none of the current models in literature on the extra-gravitational energy is able to justify alone the evidences we obtained on the entropy, metallicity and gas+dark matter temperature profiles.
\end{abstract}

\begin{keywords}
galaxies: cluster: general -- X-ray: galaxies -- intergalactic medium --
cosmology: observations.
\end{keywords}

\section{INTRODUCTION}\label{intro}
Clusters of galaxies are the biggest virialized structures in the universe, that form at a relatively late time ($z \la 2$). In the hierarchical scenario, the cosmic structures evolve hierarchically from the primordial density fluctuations, that are amplified and then collapse and merge to form larger systems under the action of gravity. The cosmic baryons fall into the gravitational potential of the cluster dark matter halo formed in this way, while the collapse and subsequent shocks heat up the intra-cluster medium (ICM) to the virial temperature ($0.5 \la T \la 10$ keV).
In the adiabatic scenario, the gravity is the only responsible of the physical properties of the galaxy clusters, once they are rescaled with respect to their masses and epoch of formation (\cite{evrard1991}).
X-ray properties of galaxy clusters show, however, some deviations form this scenario, breaking up the self-similarity predicted by the adiabatic model (see recent work in \cite{arnaud2005,donahue2006,ponman2003} and reference therein).
In particular, in the last years, the studies about X-ray scaling relations (\cite{ettori2004b}; \cite{vikhlinin2005}; \cite{kotov2005}; \cite{maughan2006}) and observations of the entropy profiles (\cite{ponman1999}; \cite{ponman2003}) in groups and clusters of galaxies, and the analysis of simulated sources with an extra non-gravitational energy injection (\cite{Borgani2004b}) have suggested that we have to account for further non-gravitational feedback beyond the gravitational energy.

The gas entropy records the thermodynamic history of the ICM as the product of both gravitational and non-gravitational processes, shaping its observed structure accordingly (\cite{voit2005a}).
The measurements of the gas entropy at 0.1 $R_{200}$ (hereafter $S_{0.1}$) showed that the observed value of $S$ is higher than the expected one from the adiabatic scenario \citep{ponman1999,Lloyd-Davies2000}, where $S$ should scale simply with the mean temperature of the virialized systems. Instead, an excess in the entropy, with respect to the prediction of the adiabatic model, is observed in the inner regions of groups and poor clusters at some fraction of $R_{200}$. This excess sets a minimum value of the entropy, labeled as entropy ``floor'' or ``ramp'', associated to the ambient gas. The presence of this minimum level of entropy calls for some energetic mechanism, not referable to the gravity only, that falls into three main classes: preheating, where the gas collapsing into the dark matter potential well is preheated by some sources, before clusters were assembled at an early epoch \citep{kaiser1991,balogh1999,tozzi2001,Borgani2005}; local heating by, e.g., AGN activity, star formation or supernovae \citep{bialek2001,brighenti2006,babul2002,borgani2002}; cooling, which seems to be able to remove low-entropy gas in the centre of the clusters, producing a similar effect to non-gravitational heating \citep{bryan2000,muanwong2002,Borgani2004b}.

In the present work, we aim to confront the model of preheating, feedback and cooling with the observed properties of the gas and of the dark matter in X-ray luminous galaxy clusters, by putting constraints on the sources of non-gravitational heating. To do that, we have considered the sample of 24 clusters presented in \cite{morandi2007}, hereafter Paper I.
To quantify the excess of energy stored in the ICM with respect to the amount available from the gravitational potential, we compare the gas and dark matter temperature profile and measure the energy feedback as a function of the radial distance. Moreover, the clusters in our sample span a wide range of redshift ($0.14 \le z \le 0.82$) and have different state of relaxation. We can thus investigate the dependence of the extra-gravitational energy feedback on the cosmic time of differently evolved structures.

This paper is constructed in this way: in Sect.~\ref{dataan}, we describe the X-ray data reduction and analysis; in Sect.~\ref{scales2}, we present our results about the entropy distribution in our hot ($kT_{\rm gas} > 6$ keV) clusters, studying its relation with the gas temperature and metallicity, its radial profile and how it relates to excess of energy measured by comparing gas and dark matter temperatures.
We discuss our results in Sect.~\ref{snnen} and summarize our findings in Sect.~\ref{conclusion}.

Hereafter we have assumed a flat $\Lambda CDM$ cosmology, with matter
density parameter $\Omega_{0m}=0.3$, cosmological constant density
parameter $\Omega_\Lambda=0.7$, and Hubble constant $H_{0}=70 \,{\rm
km/s/Mpc}$. Unless otherwise stated, we estimated the errors at the
68.3 per cent confidence level. 

\begin{figure*}
\hbox{
\psfig{figure=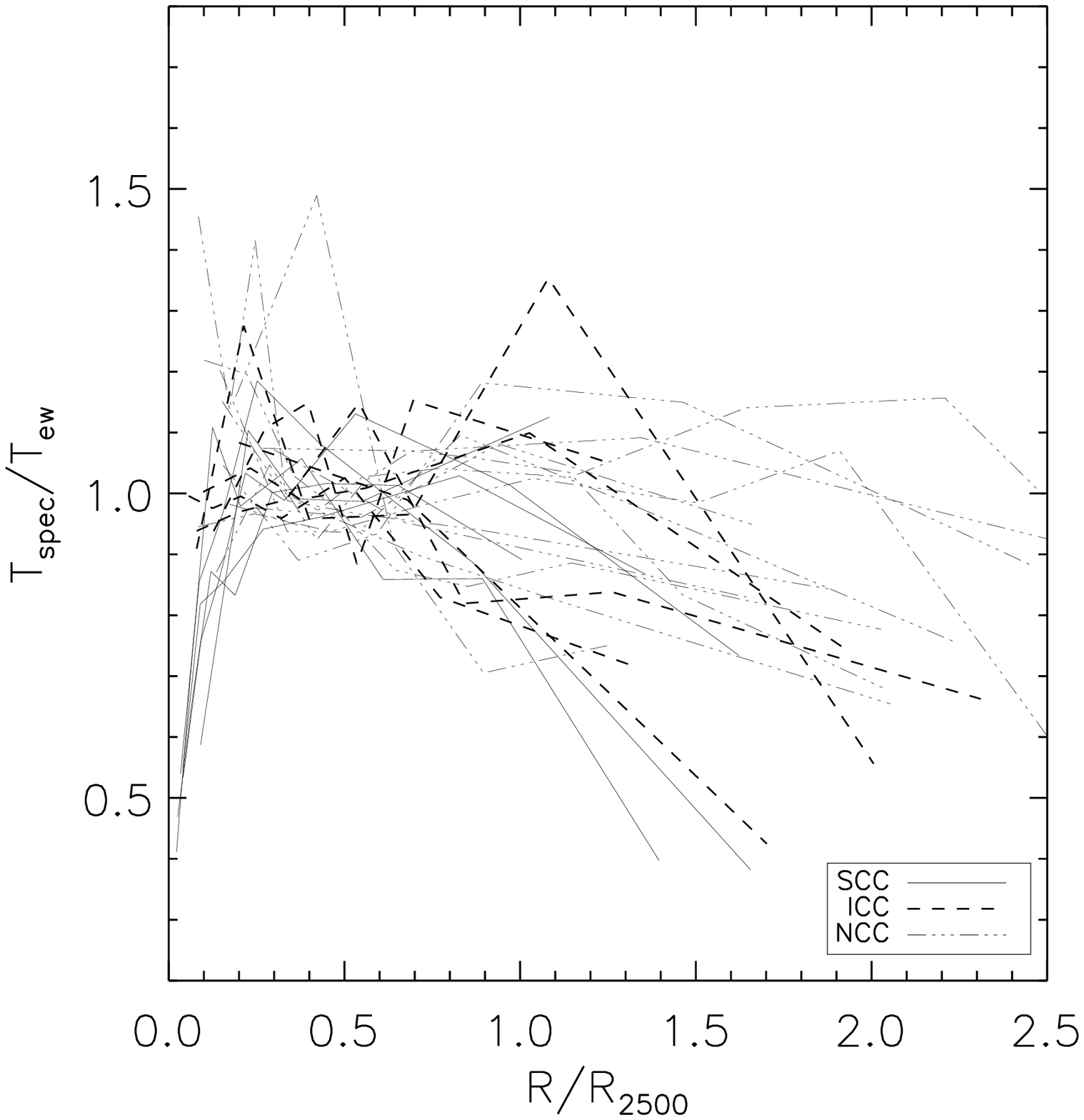,width=0.35\textwidth}
\psfig{figure=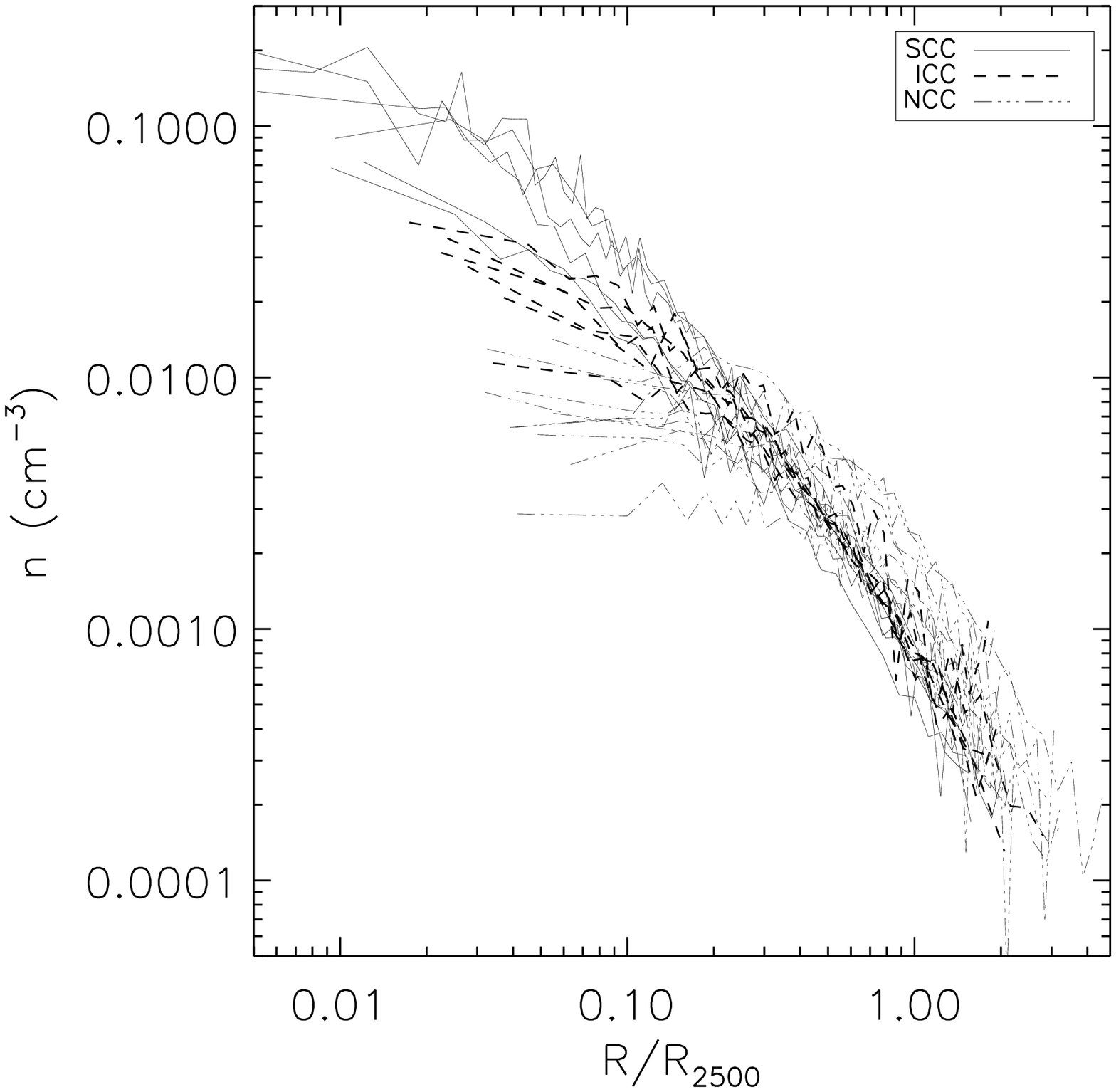,width=0.35\textwidth}
\psfig{figure=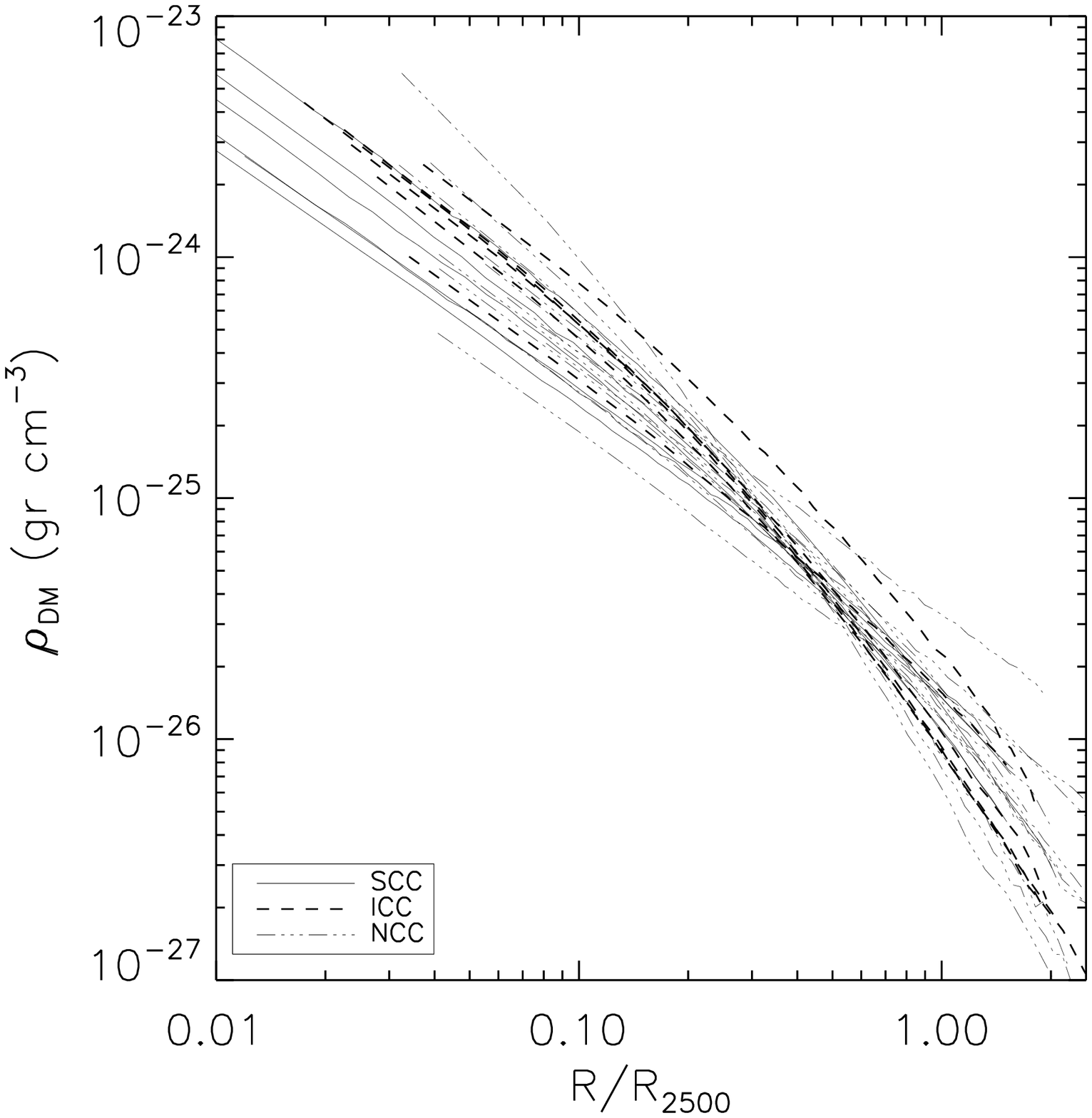,width=0.35\textwidth}
}
\caption[]{The radial profiles for the projected temperature $T_{\rm spec}(r)$,
normalized using the cooling-core corrected temperature $T_{\rm ew}$, for the gas and dark matter density are shown for all objects of our sample in the left, central and right panels, respectively. The dashed lines refer to the intermediate cooling core clusters (ICC), the solid to the strong cooling clusters (SCC), and the dot-dashed to the non-cooling core clusters.}
\label{entps3}
\end{figure*}

\begin{figure*}
\hbox{
\psfig{figure=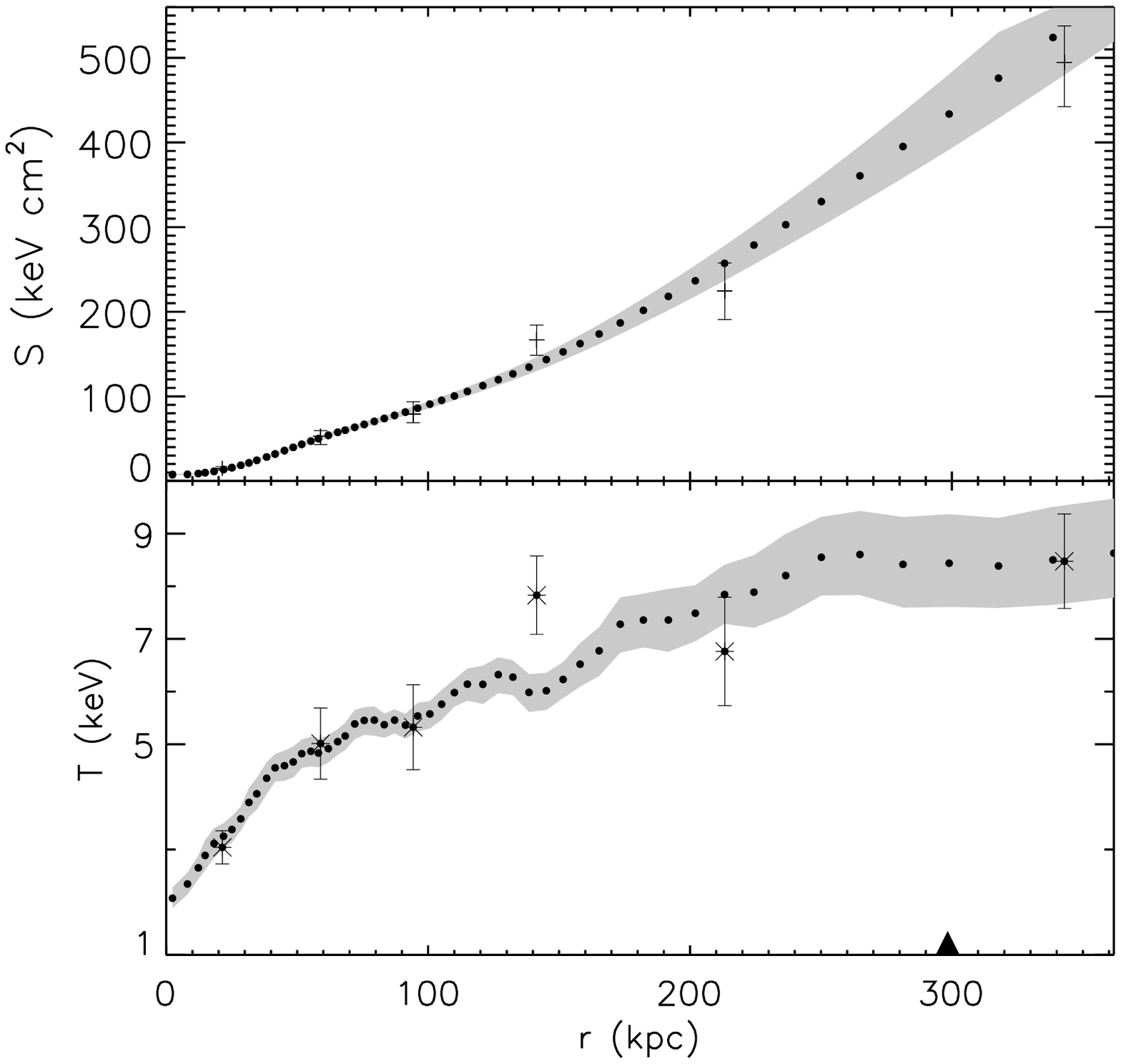,width=0.35\textwidth}
\psfig{figure=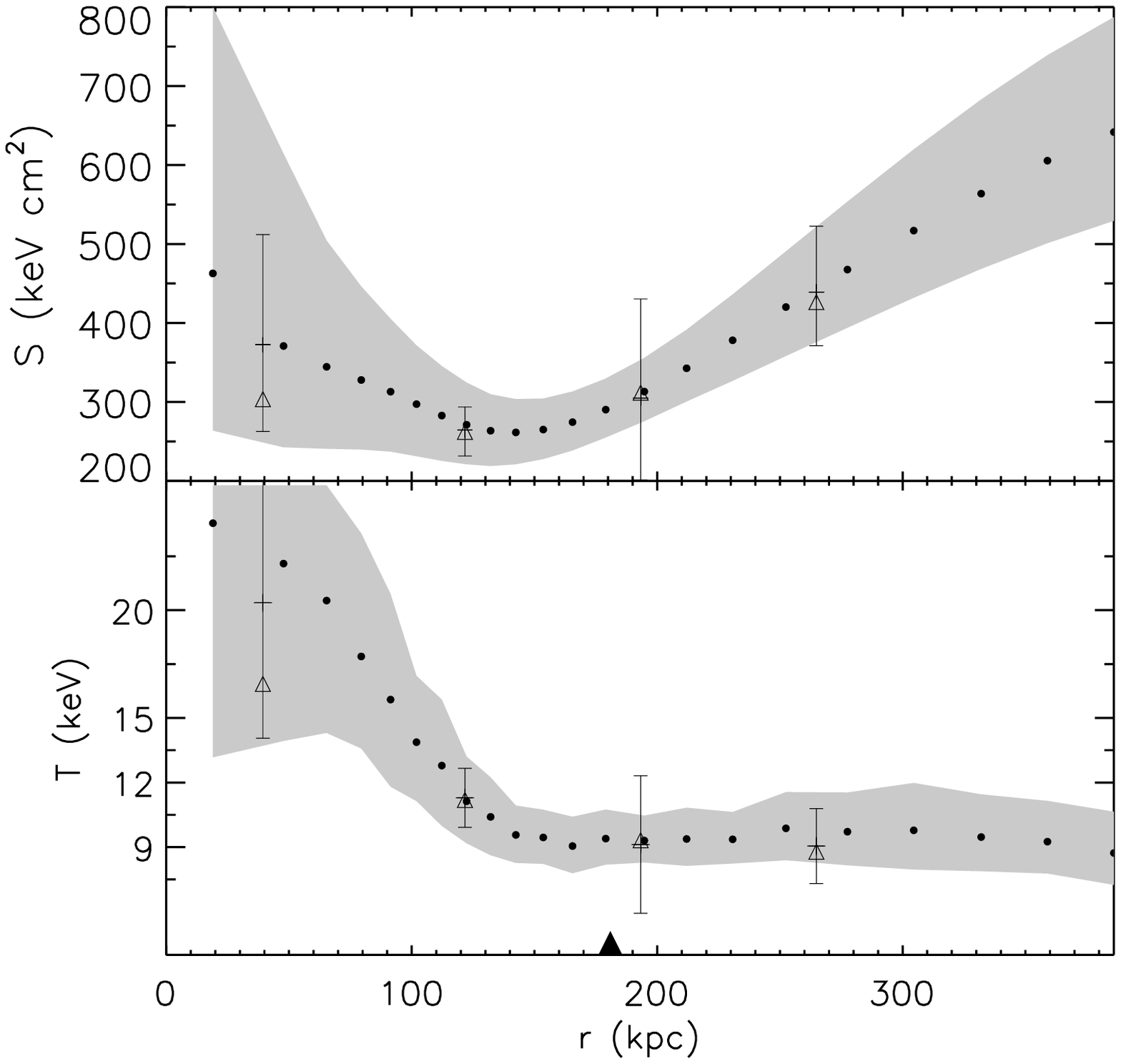,width=0.35\textwidth}
\psfig{figure=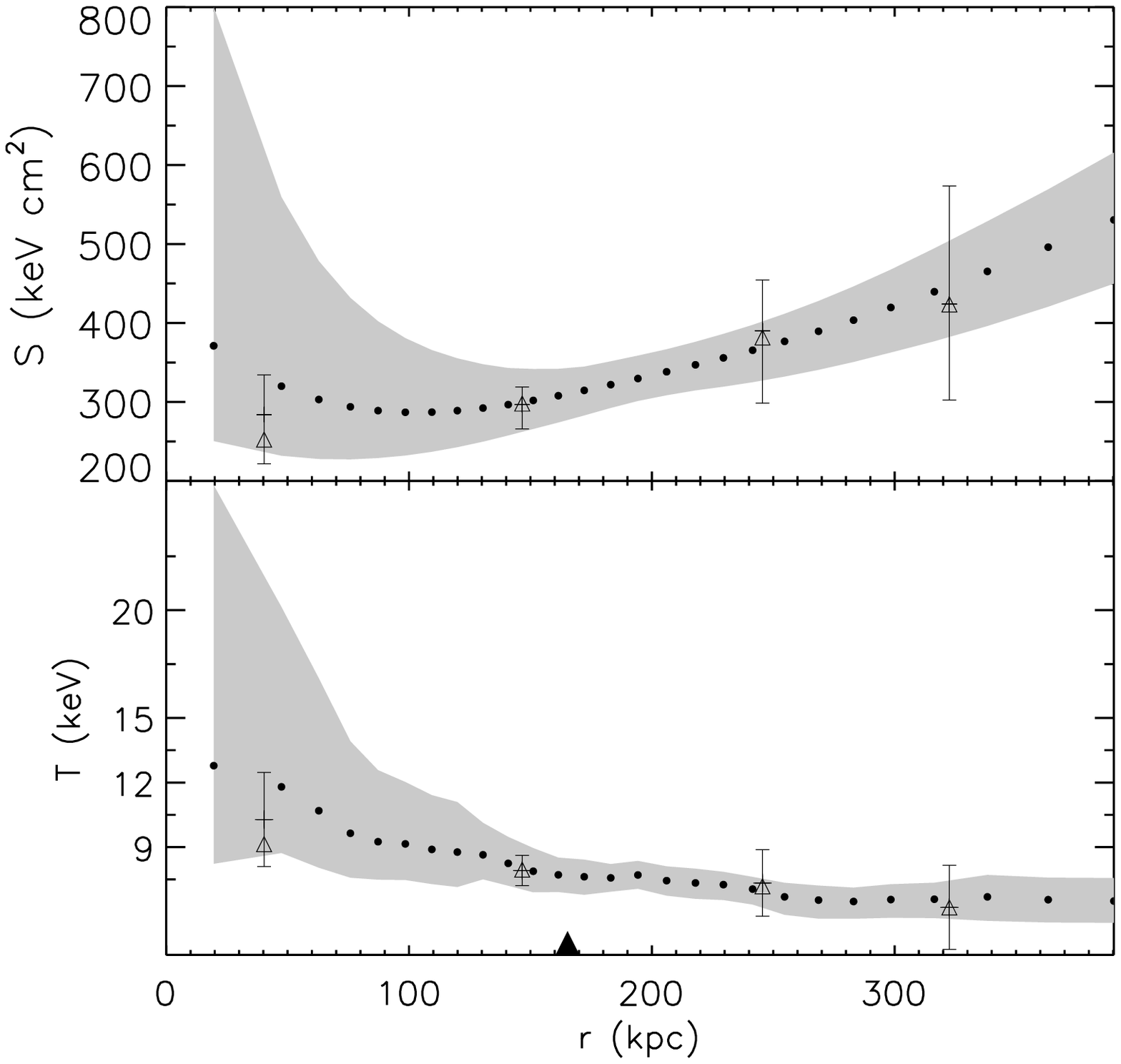,width=0.35\textwidth}
}
\caption[]{Comparison of the entropy and temperature profiles in the internal regions for, from the left to the right,
ZW3146 (SCC), A1914 (NCC) and A2218 (NCC).
The points represent each of the measure of $S_j$ in the $j$-th spherical shell by applying the analysis described in
Sect.~\ref{dataan}, while the gray region refers to the $1-\sigma$ error band. The points with errorbars (triangles)
are the measure of $S$ by applying the spectral analysis (see Sect.~\ref{entdpr}) with (without) applying the spectroscopic-like temperature definition of \citep{mazzotta2004}.
 The two triangles on the $x$-axis refer to the $\delta=0.1$ and $\delta=0.3$ (see Sect. \ref{scales2}).}
\label{entps23bb}
\end{figure*}

\begin{table*}
\begin{center}
\caption{Properties of the sample analyzed. 
For each object, the name, the redshift $z$, the emission-weighted temperature $T_{\rm ew}$,
the ratio $t_{\rm cool}/t_{\rm age}$, a flag for the presence of a strong cooling core,
an intermediate one or absence of a cooling core (labeled SCC, ICC and NCC, respectively)
are indicated.
The columns 6 and 7 refer to the best fit parameters $S_{0.1}$ and $\alpha$ for the eq. \ref{eq1} by setting $S_0 = 0$. The last four columns refer to the best fit parameters $S_{0}$, $S_{0.1}$, $\alpha$ and the total $\chi^2$ with the number of degree of freedom (d.o.f.) for the eq. \ref{eq1}. For the redshift and for each of the best fit parameters we report the average value at the bottom, by grouping the sources into SCC, ICC and NCC clusters, while the relative errors refer to dispersion of the average values.}
\begin{tabular}{l@{\hspace{.7em}} c@{\hspace{.7em}} c@{\hspace{.7em}} c@{\hspace{.7em}}
c@{\hspace{1.7em}} 
c@{\hspace{.7em}} c@{\hspace{1.7em}} 
c@{\hspace{.7em}} c@{\hspace{.7em}} c@{\hspace{.7em}} c@{\hspace{.7em}} }
\hline \\ 
& & & & & \multicolumn{2}{c}{power law} & \multicolumn{3}{c}{power law + $S_{0}$} & \\
\hline \\ 
source& $z$ & $T_{\rm ew}$ & $t_{\rm cool}/t_{\rm age}$ & CC & $S_{0.1}$ & $\alpha$ & $S_{0}$  & $S_{0.1}$ & $\alpha$ & $\chi^2{\rm d.o.f.}$  \\ 
 &  & & & /NCC & ${\rm keV\, cm^{2}}$ &   & ${\rm keV\, cm^{2}}$  & ${\rm keV\, cm^{2}}$ &  & \\ 
\hline \\
A2204 & 0.152 & $ 9.18_{-0.65}^{+0.75}$ & 0.03 & SCC & 602.6$ \pm 10.0$ & 1.05$ \pm 0.04$ & 5.8$ \pm 0.8$ & 602.4$ \pm 20.7$  & 1.44$ \pm 0.03$ & 35.9(31) \\
A2390 & 0.232 & $10.18_{-0.21}^{+0.23}$ & 0.06 & SCC & 524.8$ \pm 4.7$  & 1.30$ \pm 0.02$ & 3.8$ \pm 1.3$ & 597.6$ \pm 5.0$  & 1.15$ \pm 0.01$ & 377.6(55) \\
A1835 & 0.253 & $ 8.62_{-0.54}^{+0.60}$ & 0.09 & SCC & 288.4$ \pm 2.7$  & 1.16$ \pm 0.02$ & 11.9$ \pm 2.5$ & 273.0$ \pm 10.9$  & 1.35$ \pm 0.06$ & 10.2(30) \\
Zw3146 & 0.291 & $ 7.35_{-0.26}^{+0.27}$ & 0.02 & SCC & 457.1$ \pm 9.2$  & 1.20$ \pm 0.06$ & 4.8$ \pm 0.4$ & 401.1$ \pm 7.5$  & 1.39$ \pm 0.02$ & 55.3(57) \\
MS1358.4+6245 & 0.327 & $ 7.51_{-0.61}^{+0.70}$ & 0.05 & SCC & 489.8$ \pm 3.1$  & 1.05$ \pm 0.02$ & 0.0$ \pm 0.0$ & 511.8$ \pm 24.2$  & 1.19$ \pm 0.04$ & 9.9(31) \\
RXJ1347.5-1145 & 0.451 & $13.92_{-0.93}^{+1.14}$ & 0.08 & SCC & 616.6$ \pm 13.2$  & 1.07$ \pm 0.04$ & 15.7$ \pm 1.2$ & 548.3$ \pm 14.8$  & 1.44$ \pm 0.03$ & 22.3(44) \\
\hline \\
&0.284 &&&&  409.9$ \pm 152.6$  & 1.18$ \pm 0.11$  & 6.0$ \pm 5.9$ & 509.0$ \pm 130.7$  & 1.23$ \pm 0.16$ &  \\\hline \\
A1413 & 0.143 & $ 6.25_{-0.33}^{+0.36}$ & 0.22  & ICC & 263.0$ \pm 2.3$  & 0.95$ \pm 0.01$  & 34.8$ \pm 21.2$ & 230.9$ \pm 27.8$  & 1.06$ \pm 0.15$ & 0.8(16) \\
A1689 & 0.183 & $ 8.72_{-0.56}^{+0.63}$ & 0.18  & ICC & 316.2$ \pm 3.7$  & 1.02$ \pm 0.02$  & 36.4$ \pm 20.4$ & 286.9$ \pm 27.1$  & 1.09$ \pm 0.13$ & 1.8(23) \\
A2261 & 0.224 & $ 7.47_{-0.47}^{+0.53}$ & 0.27  & ICC & 263.0$ \pm 1.5$  & 1.09$ \pm 0.01$  & 55.1$ \pm 13.1$ & 205.3$ \pm 18.4$  & 1.29$ \pm 0.13$ & 0.8(22) \\
A611 & 0.288 & $ 6.06_{-0.34}^{+0.38}$  & 0.13 & ICC & 229.1$ \pm 0.7$  & 1.08$ \pm 0.01$  & 36.8$ \pm 14.4$ & 196.5$ \pm 19.5$  & 1.16$ \pm 0.14$ & 0.7(18) \\
A1995 & 0.319 & $7.56_{-0.41}^{+0.45}$  & 0.92 & ICC & 331.1$ \pm 3.8$  & 1.36$ \pm 0.02$  & 146.8$ \pm 22.5$ & 205.6$ \pm 34.6$  & 1.78$ \pm 0.27$ & 0.4(27) \\
MS1137.5+6625 & 0.784 & $ 5.48_{-0.71}^{+0.89}$ & 0.66   & ICC & 173.8$ \pm 5.1$  & 1.22$ \pm 0.04$  & 116.9$ \pm 36.6$ & 74.4$ \pm 42.4$  & 1.91$ \pm 0.67$ & 0.1(22) \\
\hline \\
&0.324 &&&&   241.3$ \pm 62.2$  & 1.07$ \pm 0.16$  & 57.8$ \pm 50.7$ & 210.3$ \pm 70.7$  & 1.20$ \pm 0.42$ &  \\\hline \\
A1914 & 0.171 & $ 8.93_{-0.45}^{+0.48}$  & 2.29 &  NCC & 302.0$ \pm 5.5$  & 0.96$ \pm 0.03$  & 230.1$ \pm 41.6$ & 82.2$ \pm 44.4$  & 1.99$ \pm 0.55$ & 6.7(23) \\
A2218 & 0.176 & $ 6.88_{-0.30}^{+0.33}$   & 2.01 &  NCC & 288.4$ \pm 5.6$  & 0.71$ \pm 0.03$  & 269.7$ \pm 33.4$ & 43.2$ \pm 28.4$  & 2.08$ \pm 0.65$ & 0.3(27) \\
A665 & 0.182 & $7.14_{-0.31}^{+0.33}$  & 1.47 &  NCC & 275.4$ \pm 1.3$  & 1.16$ \pm 0.01$  & 184.1$ \pm 19.8$ & 108.0$ \pm 20.9$  & 1.94$ \pm 0.21$ & 9.0(30) \\
A520 & 0.199 & $ 8.24_{-0.28}^{+0.31}$  & 5.35 &  NCC & 363.1$ \pm 14.0$  & 1.10$ \pm 0.06$  & 440.7$ \pm 12.1$ & 23.9$ \pm 6.5$  & 3.47$ \pm 0.28$ & 10.2(44) \\
A2163 & 0.203 & $12.00_{-0.26}^{+0.28}$  & 5.37 &  NCC & 831.8$ \pm 9.4$  & 1.00$ \pm 0.02$  & 244.2$ \pm 11.6$ & 498.3$ \pm 16.5$  & 1.62$ \pm 0.05$ & 94.6(52) \\
A773 & 0.217 & $ 7.23_{-0.52}^{+0.62}$  & 1.19 &  NCC & 281.8$ \pm 0.8$  & 0.85$ \pm 0.01$  & 176.7$ \pm 56.1$ & 114.4$ \pm 57.6$  & 1.48$ \pm 0.48$ & 0.7(41) \\
A697 & 0.282 & $10.21_{-0.75}^{+0.83}$ & 1.01  &  NCC & 371.5$ \pm 3.4$  & 0.99$ \pm 0.02$  & 184.4$ \pm 99.2$ & 198.4$ \pm 107.8$  & 1.45$ \pm 0.54$ & 0.1(28) \\
A370 & 0.375 & $ 7.37_{-0.53}^{+0.58}$ & 6.63 &  NCC & 398.1$ \pm 3.4$  & 0.38$ \pm 0.01$  & 396.4$ \pm 76.5$ & 25.4$ \pm 65.4$  & 2.06$ \pm 2.54$ & 0.7(20) \\
RXJ2228+2037 & 0.421 & $ 6.86_{-0.71}^{+0.89}$ & 2.05  &  NCC & 234.4$ \pm 2.0$  & 0.89$ \pm 0.01$  & 158.9$ \pm 111.4$ & 91.2$ \pm 110.0$  & 1.56$ \pm 1.10$ & 0.5(23) \\
MS0015.9+1609 & 0.546 & $ 8.29_{-0.43}^{+0.49}$ & 1.13  &  NCC & 245.5$ \pm 8.3$  & 1.02$ \pm 0.05$  & 173.7$ \pm 37.5$ & 91.2$ \pm 39.2$  & 1.76$ \pm 0.47$ & 0.3(14) \\
MS0451.6-0305 & 0.550 & $ 9.09_{-0.61}^{+0.70}$  & 0.97  &  NCC & 229.1$ \pm 4.8$  & 1.01$ \pm 0.04$  & 177.5$ \pm 56.2$ & 70.7$ \pm 65.2$  & 1.95$ \pm 1.01$ & 0.1(14) \\
EMSS1054.5-0321 & 0.823 & $ 9.00_{-1.10}^{+1.39}$ & 2.82 &  NCC & 398.1$ \pm 2.1$  & 1.00$ \pm 0.02$  & 347.2$ \pm 107.5$ & 69.2$ \pm 133.2$  & 3.01$ \pm 3.54$ & 0.1(19) \\
\hline \\
&0.345 &&&&  292.6$ \pm 174.3$  & 0.95$ \pm 0.21$  & 300.7$ \pm 110.3$ & 86.8$ \pm 132.5$  & 1.69$ \pm 0.71$ &  \\\hline \\
\end{tabular}

\label{tabdon}
\end{center}
\end{table*}

\section{The dataset and the analysis}\label{dataan}

In Paper I, we describe our dataset and the analysis applied to study
their X-ray and Sunyaev-Zel'dovich properties. Here, we remind the main characteristics of the 
sample and of the X-ray analysis adopted to recover the radial distribution of the ICM entropy
investigated in the present work.

We consider 24 galaxy clusters in the redshift range 0.14--0.82, emission-weighted temperature between 6 and 12 keV and X-ray bolometric luminosity $L \gesssim 10^{45}$ erg/s, with exposures available in Chandra archive. 
Assuming a spherically symmetric emission, 
the electron density and temperature profiles are obtained by deprojecting 
both the surface brightness profile put in hydrostatic equilibrium with a functional
form of the dark matter (DM) profile and the best-fit results obtained in the 
spatially-resolved X-ray spectral analysis by fitting a single thermal component.
In particular, from the surface brightness profile resolved in a number
of radial bins between 24 and 239, we obtain directly from the geometrical deprojection
the electron density $n_j$ in each $j$-th spherical shell.
The deprojected gas temperature, $T_j({\bf q},P_0)$, is obtained by integration of the 
hydrostatic equilibrium equation once a functional form of the dark matter density profile,
$\rho=\rho({\bf {r, q}})$, is assumed, where ${\bf q}=$(scale radius, concentration parameter) 
and the gas pressure $P_0$ at the X-ray boundary $R_{\rm spec}$ are free parameters. 
To parameterize the cluster mass distribution, we have considered two DM models: 
the universal density profile proposed by \cite{navarro1997} (hereafter NFW) and 
the one suggested by \cite{rasia2004} (hereafter RTM). 
In this study, we adopt the RTM model. Our results are not affected if a NFW functional form is used.
To constrain the 3 free parameters $({\bf q},P_0)$, we define a grid of values and proceed with a $\chi^2$ minimization of the merit function that compares the observed temperature profile with the projection of $T_j({\bf q},P_0)$ by applying the spectroscopic-like temperature definition \citep{mazzotta2004}. The best-fit values of $({\bf q},P_0)$ are the ones corresponding to the minimum $\chi^2$, $\chi^2_{\rm min}$. The associated errors are estimated at the 68.3 per cent confidence level and are computed by looking to the regions in the parameter space where $\Delta \chi^2 = \chi^2 - \chi^2_{\rm min}$ is smaller than a given threshold, 
fixed according to the number
of degrees of freedom \citep[e.g., $\Delta \chi^2 = 1, 2.3, 3.53$ for 1, 2 and 3 d.o.f., respectively;
see Paper I and ][]{press1992}. The value of $({\bf q},P_0)$ and the related errors are quoted in Paper I.

Furthermore, we deproject the best-fit results of the X-ray spectral analysis, spatially resolved in
a lower number of bins (between 4 and 10) than the surface brightness profile as requested
from the higher counts statistic needed to constrain adequately the measurements of the
temperature. However, in each $k$-th shell, the electron density $n_{k}$ and temperature $T_{k}$ 
are then recovered without any assumption of the hydrostatic equilibrium
and provide a direct verification of the validity of this assumption once they are compared
to the measures of $n_j$ and $T_j$ described above.
The spectral deprojection of the observed projected temperature $T_{\rm proj}$
has been performed in a set of $n$ annuli selected to collect
at least 2000 net counts by inverting the following equation:
\begin{equation}\label{aa1}
T_{\rm proj} = {\left({\mathcal{V}}
\# {\left( {T_k} {{n_k^2 T_k^{-\alpha}}} \right)} \right)} \; / \,
{\left({\mathcal{V}} \# {\left( {{n_k^2 T_k^{-\alpha}}} \right)} \right)},
\end{equation}
where the operator $\#$ indicates the matrix product (rows by columns), $\mathcal{V}$ is the effective volume described in Appendix of Paper I, and $\alpha=0.75$ using the spectroscopic-like temperature definition \citep{mazzotta2004}.

\subsection{Cooling core and Non-cooling core clusters}

In the following analysis, we divide our sample in three categories, depending
on the strength of the central cooling-core  (see Table~\ref{tabdon}):

\begin{itemize}
\item {\bf Strong cooling core (SCC)} clusters are the 6 objects in which
the central cooling time is significantly less than the age of the universe
at the cluster redshift ($t_{\rm cool}/t_{\rm age, z} <0.1$). 
They show very low central temperature ($\sim 2$ keV) and strong spike of 
luminosity in the brightness profile, and a very pronounced drop of the temperature 
near the boundary of the observation, about a factor two compared with the peak of the temperature. The temperature profile is very regular, suggesting a relaxed
dynamical state.

\item {\bf Intermediate cooling core (ICC)} clusters have a central cooling time with values
$0.1 \lesssim t_{\rm cool}/t_{\rm age, z} \lesssim 1$. The six objects in our sample
show a less prominent spike of brightness than SCC clusters and a mild drop of the temperature
in the cooling region ($\gesssim 1/2 \, T_{\rm ew}$).

\item The {\bf Non-cooling core (NCC)} sources (12 objects in our sample)
have central cooling time higher than $t_{\rm age, z}$ and do not present any evidence of the 
central drop in the temperature profile. Both the temperature profile and surface brightness map
are less regular than the ones observed in CC systems, showing hints of substructures
and merging activity.
\end{itemize}

The gas and DM density profiles (right panel of Figure~\ref{entps3}) have similar slopes
over the entire radial range in the SCC clusters, whereas less 
self-similarity is present in the ICC and especially in the NCC clusters:
the gas density profile is here flatter than the $\rho_{\rm DM}$ one,
supporting the scenario in which the ICM has been affected by some form 
of non-gravitational energy.
We discuss the physical interpretation of these observational results in Section~\ref{snnen0}.

The high level of relaxation of the SCC sources is also confirmed by the study of the polytrophic index $\gamma$
\footnote{$\gamma$ is calculated as $\equiv {d\, \log{(T_k)}}/ {d\,\log{(n_k)}} +1$ by linear fit in the
$\log{(n_k)} - \log{(T_k)}$ plane by considering the spectral deprojected density $n_k$ and temperature $T_k$ 
described in Sect.~\ref{dataan}.}, that has values near 1 with a very low scatter for the SCC sources, 
whereas is more scattered in NCC sources at $r\gesssim 0.5R_{2500}$:
$\gamma_{\rm SCC}=1.01 \pm0.09$, $\gamma_{\rm ICC}=1.06 \pm0.12$, $\gamma_{\rm NCC}=1.08 \pm0.32$.
Within $0.3 \, R_{200}$, we measure $\gamma_{\rm SCC}=0.66\pm0.07$, $\gamma_{\rm ICC}=0.97\pm0.05$ and 
$\gamma_{\rm NCC}=1.29\pm0.50$, with a clear increase as a function of the morphological type
and a very high scatter for the NCC sources.

\subsection{On the gas entropy profile}\label{entdpr}

In the present paper, we have extended the above analysis by estimating the entropy profile
in each cluster by using (i) the gas pressure $P_j $ and density $n_j$ profile in the equation 
$S_j=P_j/n_j^{5/3}$ and (ii) the deprojected spectral results $S_{k}=T_k/n_k^{2/3}$. 
The errors on the entropy profiles are obtained by error propagation of the uncertainties on the
single quantity and/or best-fit parameters. 
We note that the dependence of $S_j$ over $P_0$, the gas pressure value
at the X-ray spectral boundary, can be checked by comparing it with the entropy measured once $P_0$
is fixed to the value measured in the spectral analysis: 
we find a totally negligible variation at $0.1\, R_{200}$ and a change $\lesssim 5$ per cent at $0.3\, R_{200}$.

In Figure~\ref{entps23bb}, we present a comparison of the entropy and temperature profiles recovered
with the two methods in the inner regions of three representative cases,
ZW3146 (SCC), A1914 (NCC) and A2218 (NCC). 
We obtain good agreement between the entropy measurements in ZW3146 and A1914, 
the former being an example of a typical CC source where 
the profile decreases moving inward, whereas the latter shows the most evident case
of flattening, with hints of inversion, in the central entropy values.
Given the good agreement between $S_{k}$ and $S_j$ even in the internal regions, 
we believe that this inversion is not due to our approach, but it is real in A1914
(similar behaviour is found in A773).
We note that, if we use the entropy recovered by using the proper cooling function 
in eq. \ref{aa1} instead of the functional $T^{-\alpha}$, 
this inversion is less pronounced (see Fig.~\ref{entps23bb}) for $S_{k}$.

In A2218 (NCC), we observe a marginal disagreement between $S_j$ and $S_{k}$:
$S_j$ shows an inversion in the core, whereas $S_{k}$ appears flatter. 
We draw similar conclusion for A370, A520, A2163, and RXJ2228+2037.
Nevertheless we observe that for the latter sources the low spatial resolution of $S_{k}$ in the central regions 
($\gesssim 150-200$ kpc) does not allow to sample properly $S_{j}$ on scales of $\lesssim 50-150$ kpc,
where the inversion occurs. 

We note that the larger deviations between $T_j$ and $T_k$ are observed in NCC clusters within 100 kpc,
where we expect higher relative contribution from non-thermal effects due to, e.g., merging activity.
Therefore, even though the most prominent substructures identified in the cluster images were masked, 
implying that we have reduced their effects in the temperature reconstruction under the hydrostatic 
equilibrium equation, the sampled gas might be still subjected to ongoing merging processes. 
The higher value of $S_{j}$ compared to $S_{k}$ in the cluster centre is likely due to a very 
flat density profile that induces a higher temperature value (once the hydrostatic equilibrium 
equation is applied) than the spectral deprojected temperature.
Indeed, unresolved mergers could lead to this very flat density profile
(they are clearly visible in A520 and A2163), if the gas at $R\lesssim50-100$ kpc is not wholly relaxed
and in hydrostatic equilibrium. For the other NCC sources, that do not show clearly ongoing merging processes,
nevertheless we noted a disturbed morphology, as indicated, for example, from the fact that 
the centroid of symmetry does not coincide with the peak of brightness.

On larger scales, that involve larger cluster volumes, local deviations from the hydrostatic equilibrium 
are washed out even in the most unrelaxed objects, making tenable the hypothesis upon which $S_j$ is obtained.
This is also confirmed from (i) the agreement between $S_{j}$ and $S_k$ ($T_{j}$ and $T_k$) measured 
in these sources, (ii) the results of hydrodynamical numerical simulations \citep{rasia2006}, and
(iii) the analysis presented in Paper I (Sect. 4.1.1), where we show how the relation between
$M_{\rm tot}$ and the mass-weighted temperature for our sample is in agreement with the results coming
from simulations including feedback and radiative processes, supporting our overall mass and 
temperature $T_j$ reconstruction.

In the following analysis, we evaluate the entropy at 0.1 and $0.3\,R_{200}$, i.e. at radii
well beyond the region where the central inversion of $S(r)$ is observed in few NCC objects.
Given that, and the good agreement on larger scales between the reconstructed profiles,
we define $S(r)=S_j$ hereafter to fully exploit the spatial resolution available in estimating
the entropy radial profile.

\section{Entropy and temperature distribution}\label{scales2}

We examine the $S-T$ relation at fixed overdensities, comparing our results with the ones available in literature
for nearby systems. We investigate, then, the radial entropy profile, studying the behaviour of its gradient
and its dependence upon the state of relaxation of the system. Finally, we implement an analysis 
of the temperature of the ICM and of the DM to quantify the excess of energy associated
to the gas and its radial distribution.

\subsection{The entropy-temperature relation}\label{scales1}

\begin{figure*}
\hbox{
\psfig{figure=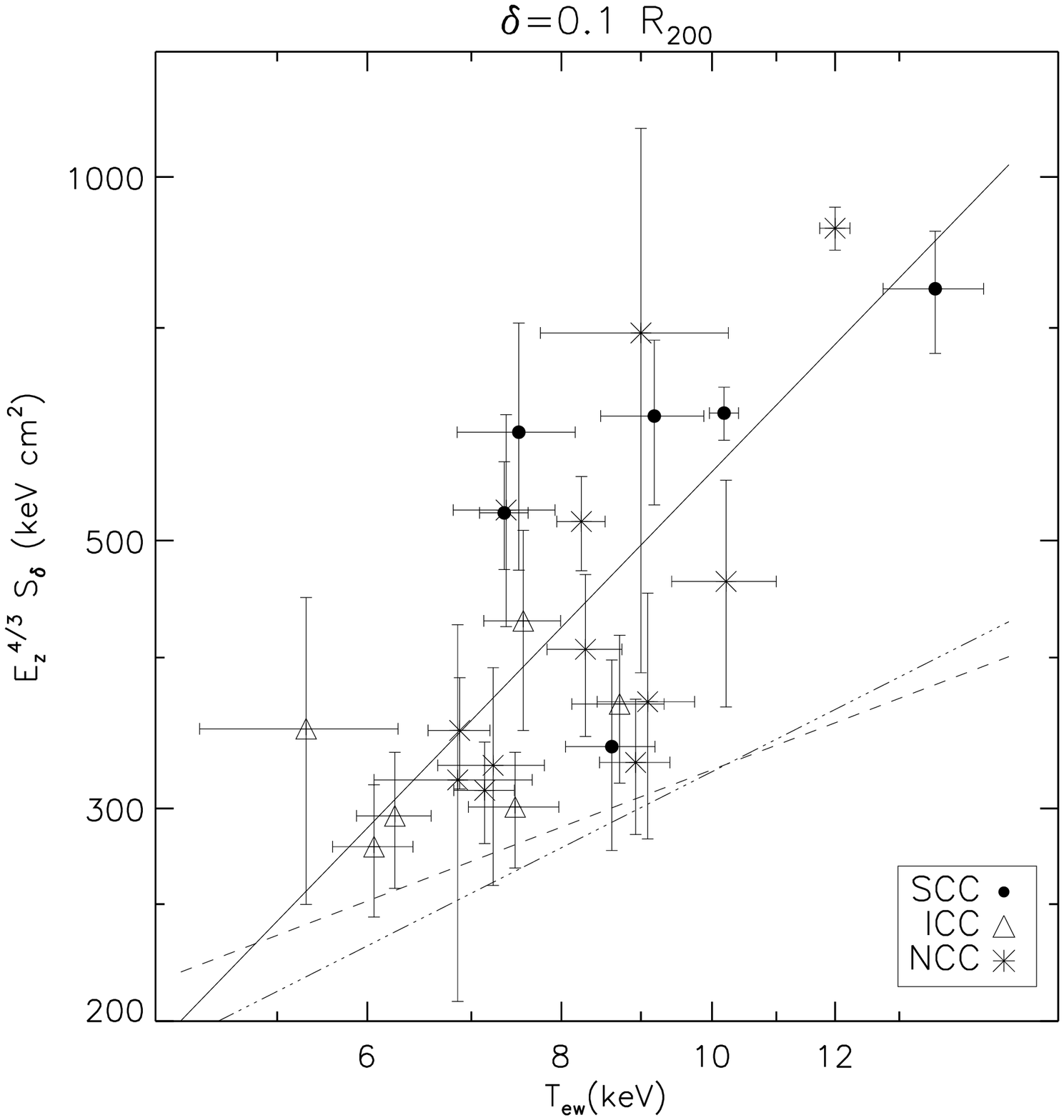,width=0.5\textwidth}
\psfig{figure=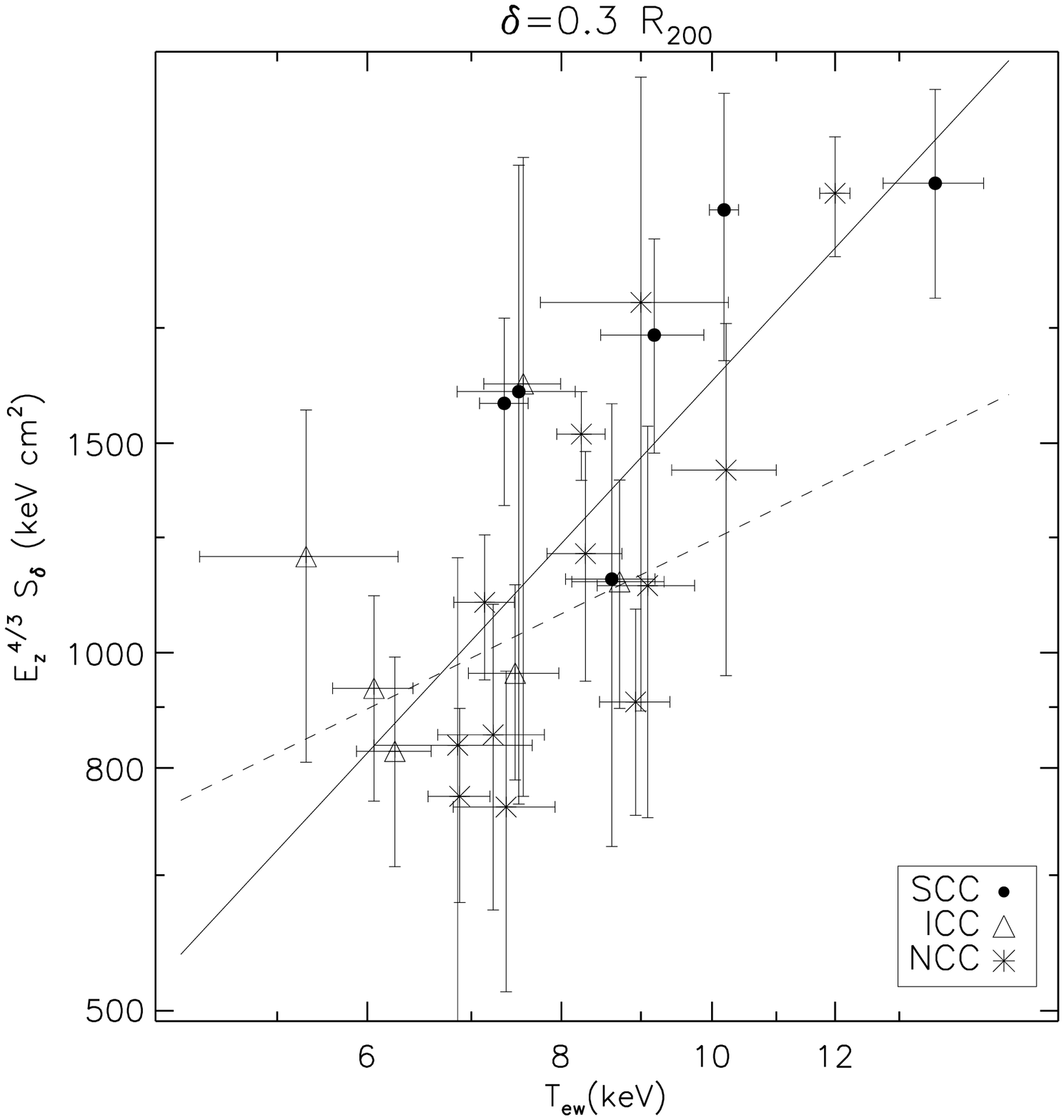,width=0.5\textwidth}
}
\caption[]{The $S-T$ relation a different fraction $\delta$ of $R_{200}$: $\delta=0.1\, R_{200}$ (left panel) and $\delta=0.3\, R_{200}$ (right panel). In each panel the filled circles represent the strong cooling core sources (SCC), the  triangles the intermediate cooling core clusters (ICC), while the stars the non-cooling core clusters.  The solid line refers to the best-fit relation obtained when considering all clusters of our sample, while the dashed one represents the best-fit obtained by \cite{pratt2006} and the dot-dashed by \cite{piffaretti2005}.}
\label{entps1a}
\end{figure*}

We have determined the entropy-temperature relation at different fraction $\delta$ of the virial radius $R_{200}$ ($\delta=0.1$ and $\delta= 0.3$).
We have fitted a power-law model of the form:
\begin{equation}\label{14}
E_z^{4/3} S_{\delta}  =\alpha T_{\rm ew,7}^{A}\ \ ,
\end{equation}
where $E_z \!=\left[\Omega_M (1+z)^3 +  (1-\Omega_M-\Omega_{\Lambda})(1+z)^2 + \Omega_{\Lambda}\right]^{1/2}$ 
and $T_{\rm ew,7}$ is the total cool-core corrected (by masking the central $r=100$ kpc region) 
emission-weighted temperature in units of 7 keV (see Paper I). 
The fit has been performed by adopting the BCES (Bivariate Correlated Errors and intrinsic Scatter) $Y \!\! \mid
\!\!X$ method \citep{akritas1996} (see Paper I for further details on this approach).
We quote our best-fit results in Table~\ref{tabscal} and show the distribution of the entropy values 
at different fractions of $R_{200}$ in Figure~\ref{entps1a}.
We note that SCC clusters show higher normalization ($\sim 440$ and $1400$ keV cm$^2$ at $\delta=0.1$ and $0.3 R_{200}$, respectively) than ICC and NCC objects, with a larger deviation in the inner region ($\delta=0.1$) which can be explained invoking different relaxation states of the clusters as discussed in Sect.~\ref{metallicity}. The best-fit slopes, within the error-bar at $1 \sigma$, are in agreement with the self-similar
prediction ($A=1$) and steeper than the slopes of $A\sim 0.5-0.6$ observed 
in local samples of galaxy groups and clusters \citep{piffaretti2005,pratt2006,ponman1999,ponman2003}.  
For comparison, we present in Table~\ref{tabscal} also the normalizations measured 
by fixing $A=1$ (self-similar expectation) and $A\sim 0.65$ \citep{ponman2003} and plot in
Figure~\ref{entps1a} the best-fit results obtained by \cite{pratt2006}
and \cite{piffaretti2005} from their analyses of relaxed groups and clusters at low redshift.
\cite{pratt2006} measure $A=0.49\pm0.15\;(\alpha=271\pm20)$ and $A=0.64\pm0.11(\alpha=990\pm55)$ 
for $\delta=0.1$ and $\delta=0.3$, respectively, with a clear departure from the self-similar expectation ($A=1$).
\cite{piffaretti2005} at $\delta=0.1$ measure $\alpha=255\pm71$ by fixing $A=0.65$.
Once these results are compared with what we measure in our sample of very massive systems,
we observe that our normalizations are on average higher by 20-60 per cent,
with slopes that are steeper and closer to the self-similar prediction 
than the values measured locally (see also \cite{ponman2003}).
This result is in agreement with the fact that we are measuring the entropy distribution
in massive clusters with cool-core corrected temperatures in the range $6-12$ keV.
These systems are definitely less affected from extra-gravitational, feedback processes that,
on the contrary, are so relevant in groups and low-mass clusters representing
the bulk in the sample of objects studied in, e.g., \cite{piffaretti2005} and \cite{ponman2003}.
Moreover, by parameterizing  the evolution in redshift using a $(1+z)^B$ dependence
(see Paper I for further details on this approach),  
we did not observe any hints of evolution of the entropy-temperature relation
within our sample. A very significant positive evolution ($B \approx 2 \pm 0.1$
but with a reduced $\chi^2$ of 5; see Fig.~\ref{entps1a})
is instead measured in the relation between the entropy
estimated at $0.1 R_{200}$ and $T_{\rm ew}$ when our CC (SCC+ICC) objects are compared
to the best-fit local results in Pratt et al. (2006).
Although the local best-fits refer to objects distributed over a wider range in temperature,
the systematic larger values measured at higher redshift is noticeable and definitely
more evident at $0.1 R_{200}$ than at $0.3 R_{200}$ where we measure $B \approx 1 \pm 0.2$
with a reduced $\chi^2$ of about 1.

\begin{table}
\begin{center}
\caption{Best fit parameters of the $S-T$ relation by applying the eq. \ref{14}. The sources are grouped into SCC, SCC+ICC and all clusters}
\begin{tabular}{l@{\hspace{1.7em}} c@{\hspace{.7em}} c@{\hspace{1.7em}} c@{\hspace{.7em}} c@{\hspace{2.7em}}  }
\hline \\ 
&  \multicolumn{2}{c}{$\delta=0.1$} & \multicolumn{2}{c}{$\delta=0.3$}\\
\hline \\ 
CC & $A$ & $\alpha$  & $A$ & $\alpha$ \\ 
/NCC & & ${\rm keV\, cm^{2}}$  &  & ${\rm keV\, cm^{2}}$  \\ 
\hline \\
SCC & $0.76\pm0.24$ &  $468\pm70$ & $0.91\pm0.19$ & $1380\pm154$\\
    & $1$ (fixed)   &  $440\pm10$ & $1$ (fixed)   & $1409\pm146$ \\
    & $0.65$ (fixed)&  $494\pm20$ & $0.65$ (fixed)& $1567\pm155$ \\ 
SCC+ICC & $1.18\pm0.16$ & $ 380\pm28 $ & $1.17\pm0.22$ & $1202\pm94$\\
    & $1$ (fixed)   &  $399\pm14$ & $1$ (fixed)   & $1186\pm85$ \\
    & $0.65$ (fixed)&  $432\pm15$ & $0.65$ (fixed)& $1242\pm89$ \\ 
all & $1.33\pm0.20$ & $ 354\pm20 $ & $1.41\pm0.26$ & $ 1023\pm77$\\
    & $1$ (fixed)   &  $415\pm11$ & $1$ (fixed)& $1140\pm50$ \\ 
    & $0.65$ (fixed)&  $460\pm14$ & $0.65$ (fixed)& $1216\pm59$ \\ 
\hline \\\\
\end{tabular}

\label{tabscal}
\end{center}
\end{table}

\subsection{Properties of the entropy profiles}\label{snnen00}
To characterize the gas entropy profile, we follow \cite{donahue2006} and fit two different models.
The first one reproduces the radial entropy profile with a power law plus a constant $S_0$:
\begin{equation}\label{eq1}
S(r) = S_0 + S_{0.1} \left( \frac {r} {0.1
\,r_{200}} \right)^{\alpha}
\end{equation}
In the second functional form, we set $S_0 = 0$, modeling the entropy profile with a pure power law.
The best fit parameters on the radial profile are determined by applying the $\chi^2$ statistic to the eq.~\ref{eq1} 
over the radial entropy profile between $0.1\,R_{200}$ and $0.3\,R_{200}$, whereas BCES(Y$|$X)
is used when $S_0$ is fixed to zero and the region within $0.1\,R_{200}$ is excluded from the fit because it is strongly affected by the cooling process.
The outermost bins of the fit are excluded by the fit, being noisy and likely affected by systematic errors due to subtraction of the noise in the data reduction (see Paper I).
Our best-fit results are quoted in Table~\ref{tabdon}.

\begin{figure}
\hbox{
\psfig{figure=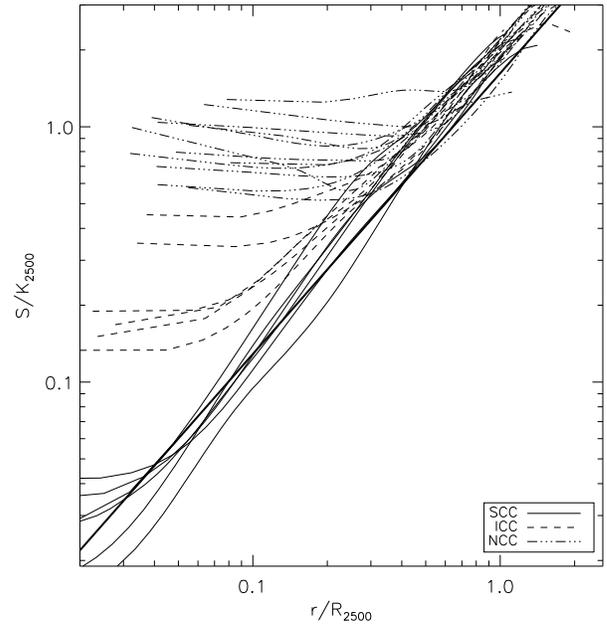 ,width=0.5\textwidth}
}
\caption[]{Profiles of $S/K_{2500}$ as a function of $R/R_{2500}$. The dashed line represents the intermediate cooling core clusters (ICC), the solid the strong cooling clusters (SCC), and the dot-dashed the non-cooling core clusters (NCC). The thick solid line represents the profile of \cite{Voit2005b}, $S/K_{2500}=1.62(r/r_{2500})^{1.1}$ (see their Fig. 1, where we have renormalized their entropy profile from $\Delta=200$ to $\Delta=2500$).}
\label{entps1}
\end{figure}

The entropy profiles show a regular behaviour (see Fig.~\ref{entps1}), once the quantities are 
rescaled to the characteristic value $K_{2500}$ at the overdensity of 
2500\footnote{$R_{2500}$ is $\sim 0.25 R_{200}$, i.e. $\approx 400-600$ kpc.} 
for adiabatic clusters (see, e.g., eq.~2 in \cite{Voit2005b}).
Profiles of CC clusters are similar down to the inner resolved regions, whereas NCC systems show large deviations in the central parts. These profiles are well reproduced by the functional form with a power-law plus a constant for which we obtain a $\chi^2_{red}$ always less than or of the order of unity, apart from A2390.
In particular, SCC sources show a very tight range of values of the entropy pedestal $S_{0}$ ($S_{0}\lesssim 15 \,\rm{keV \,cm^2}$) in agreement with the value found by Donahue et al., and a power law behaviour which is roughly preserved on the entire range of the radial entropy profile, even in the cooling region (see right panel of Figure~\ref{entps1}). The average slope determined from the second method ($\alpha=1.18 \pm0.11)$ is very similar to the theoretical value of 1.1 predicted by \cite{tozzi2001} by using analytic models of shock dominated spherical collapse. Concerning $S_{0.1}$, it shows values in the range $270-600\,\rm{keV \,cm^2}$: if we adopt the definition of $S_{100}$ in Donahue et al. as the normalization at 100 kpc, we have $S_{100} \sim 90-150\,\rm{keV \,cm^2}$, mildly lower than than the range found by them ($S_{100} \sim 90-240\,\rm{keV \,cm^2}$).

The ICC clusters show higher and wider range of $S_{0}$, with a typical value of $\sim 30\,\rm{keV \,cm^2}$. The power-law behaviour is preserved just on large scale, i.e. outside the cooling region. The average slope is still in agreement with above theoretical predictions ($\alpha=1.07\pm0.16$), but it is a little lower than the value measured in SCC clusters.

In NCC objects, we observe a more scattered radial profile, which is likely self-similar beyond the central regions ($\sim 0.5\, R_{2500} \approx 200-300$ kpc). In the inner regions, we notice a very high dispersion on the entropy pedestal value ($\sim 80-400\, \rm{keV \,cm^2}$), larger than the values found in the CC clusters. The average slope is mildly lower than the one determined in the CC-only subsample ($\alpha \sim 0.95\pm 0.21$).

We point out that $\alpha$ rises by considering NCC, ICC and SCC sources, respectively ($\alpha^{\rm SCC}=1.18\pm0.11$, $\alpha^{\rm ICC}=1.07\pm 0.16$ and $\alpha^{\rm NCC}=0.95\pm 0.21$ for the power-law model). As we will see in Sect. \ref{snnen0}, this trend is probably due to the effect of non-gravitational sources on large scale, which justify the flatter radial behaviour of the entropy profile in NCC clusters.

It is worth noticing the behaviour of the entropy pedestal $S_{0.1}$ in the different subsamples: $S_{0.1}^{\rm NCC}=300.7\pm 110.3\,\rm{keV \,cm^2}$; $S_{0.1}^{\rm ICC}=57.8\pm 50.7\,\rm{keV \,cm^2}$; $S_{0.1}^{\rm SCC}=6.0\pm 5.9\,\rm{keV \,cm^2}$. The trend of the gas density and temperature profile (see central and left panel of Figure~\ref{entps3}) can justify the progressively greater value of the entropy in the inner regions by considering SCC, ICC and NCC clusters, respectively. 
We observe higher normalization of the entropy in SCC sources 
(see Figure \ref{entps1} and the value of the parameter $A$ in Table~\ref{tabdon}). 
This behaviour is due to the fact that the SCC sources show steeper density profiles, i.e. at the same fraction of $R_{200}$, as long as we consider radii greater than $0.1\, R_{200}$, the density of the SCC sources is lower.
Even though the temperature profiles in the SCC sources are a bit steeper than in the ICC and NCC objects, the overall
effect is that the gas entropy tends to be higher in SCC clusters. We note that the unrelaxed morphology
of the NCC sources can not account for systematic changes in, e.g., the determination of $R_{\Delta}$.

On the evolution with redshift of the best-fit parameters of eq.~\ref{eq1}, we note that only for $S^{\rm ICC}_{0.1}$ we obtain a marginal evidence of negative evolution (Spearman's rank coefficient $r_{\rm s}=-0.60$ for 22 d.o.f with probability of null correlation $p=0.28$). On the contrary, $\alpha$ shows a positive evolution for the ICC clusters: $r_{\rm s}^{\rm ICC}=0.90$, with $p=0.37$, while for the SCC and NCC sources there is not apparent evolution.

We have calculated the weighted average value of the slopes of the best-fit parameters of the local sample of clusters determined by \cite{donahue2006}, so as to compare it with our estimate at higher redshift: they measure $\alpha=1.00\pm0.01$ (when $S_0=0$) and $\alpha=1.23\pm0.01$ (by accounting for $S_0$), while we obtain $\alpha=1.10\pm0.01$ and $\alpha=1.27\pm0.01$, respectively, by applying their procedure\footnote{The errors refer to the average value.}.
These results, confirmed also including in the sample the ICC sources, 
suggest that entropy profiles in nearby CC systems are slightly flatter than in CC clusters
at higher redshift, providing some marginal hints on the evolutionary trends
present in the entropy distribution.

\begin{figure}
\psfig{figure=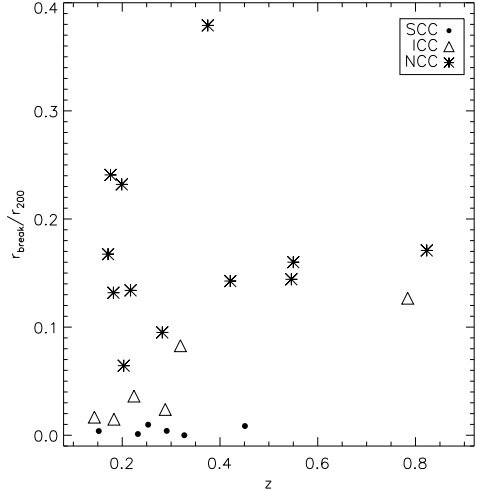,width=0.4\textwidth}
\caption[]{
Normalized break radius $r_{\rm break} /r_{200}$ as a function of the redshift.}
\label{entps23}
\end{figure}

In Figure \ref{entps23}, we plot the break radius $r_{\rm break}$ present in the entropy
profile $S(r)$, i.e. the radius where $S_0 = S_{0.1} \left( {r_{\rm break}} /{(0.1
\,r_{200})} \right)^{\alpha}$ in eq. \ref{eq1}, as a function of the redshift. 
We found the following average values for $r_{\rm break} /r_{200}$: $r^{\rm SCC}_{\rm break} /r_{200}=0.005\pm0.004$, $r^{\rm ICC}_{\rm break} /r_{200}=0.050\pm0.045$ and $r^{\rm NCC}_{\rm break} /r_{200}=0.172\pm0.082$.
The NCC sources show value of $r_{\rm break} /r_{200}$ definitely higher than the CC clusters, defining the scale where the non-gravitational energy breaks the self-similarity ($r^{\rm NCC}_{\rm break} \sim 0.1-0.4 r_{200} \sim 200-600$ kpc). We do not observe significant evolution for $r_{\rm break}$, except for the CC objects (Spearman's rank coefficient of $r_{\rm s}=0.89$ , probability of null correlation $p=0.019$).

\subsection{Gas and dark matter temperature profiles}\label{snnen0}

In this section, we define a temperature associated to the dark matter component following
the method presented in \cite{ikebe2004,2007arXiv0705.4680H}.  We define the temperature of the dark matter halo, $T_{\rm DM}$,  as: 
\begin{equation}
k T_{\rm DM} \equiv \frac{1}{3} \left( \sigma_r^2 + 2 \sigma_{\theta}^2 \right) \mu m_{\rm p}
\label{eq:temp}
\end{equation}
where $\mu$ is the mean molecular weight of the ICM, $m_{\rm p}$ is the proton mass, $\sigma_{\theta}$ and $\sigma_r$ are the 1-dimensional tangential and radial velocity dispersions of the dark matter. The radial velocity dispersions has been obtained by solving the Jeans equation:
\begin{equation}
\frac{GM({\bf q})}{R} =
- \sigma_{r}^2 \left( \frac{d \ln{\rho_{\rm DM}({\bf q})}}{d \ln{R}}
 + \frac{d \ln{\sigma_{r}^2}}{d \ln{R}} + 2 \beta({\bf q}) \right) ,
\label{eq:Jeans1}
\end{equation}
where a velocity anisotropy parameter is defined,  $\beta({\bf q})=1-\sigma^2_{\theta}/\sigma^2_{r}$.
 N-body simulations for a variety of cosmologies shows that $\beta$ has roughly an universal radial profile 
\citep[][]{cole1996}, which is given by the following relation:
\begin{equation}
\beta({\bf q})=\beta_m {{4r_n}\over{r_n^2+4}}
\label{eq:Jeans2}
\end{equation}
where $r_n=r/r_{200}({\bf q})$, and $\beta_m\approx 0.3-0.5$ \citep[][]{carlberg1997}. 
The dark matter profile is estimated as $\rho_{\rm DM}=\rho_{\rm tot} -\mu m_{\rm p} n_{\rm gas}$, 
where $\rho_{\rm tot}$ and $n_{\rm gas}$ has been determined from the analysis in Paper I.
We will compare the dark matter temperature to the gas temperature, $T_{\rm gas}$,
recovered by applying the hydrostatic equilibrium equation
\footnote{Following the notation in Sect.~\ref{dataan}, $n_{\rm gas}=n_j$ and $T_{\rm gas}=T_j$.}.

\begin{figure*}
\hbox{
\psfig{figure=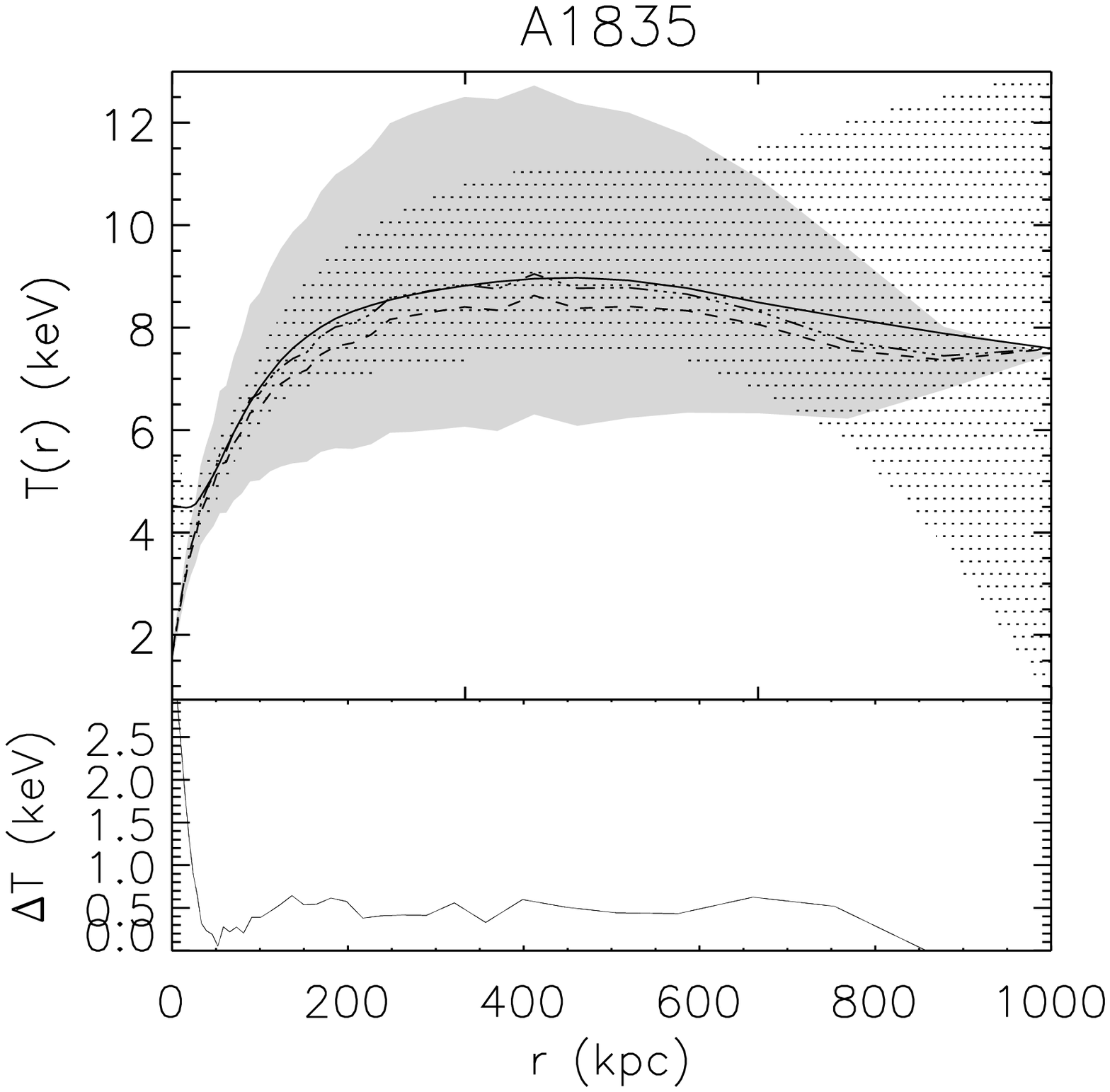,width=0.35\textwidth}
\psfig{figure=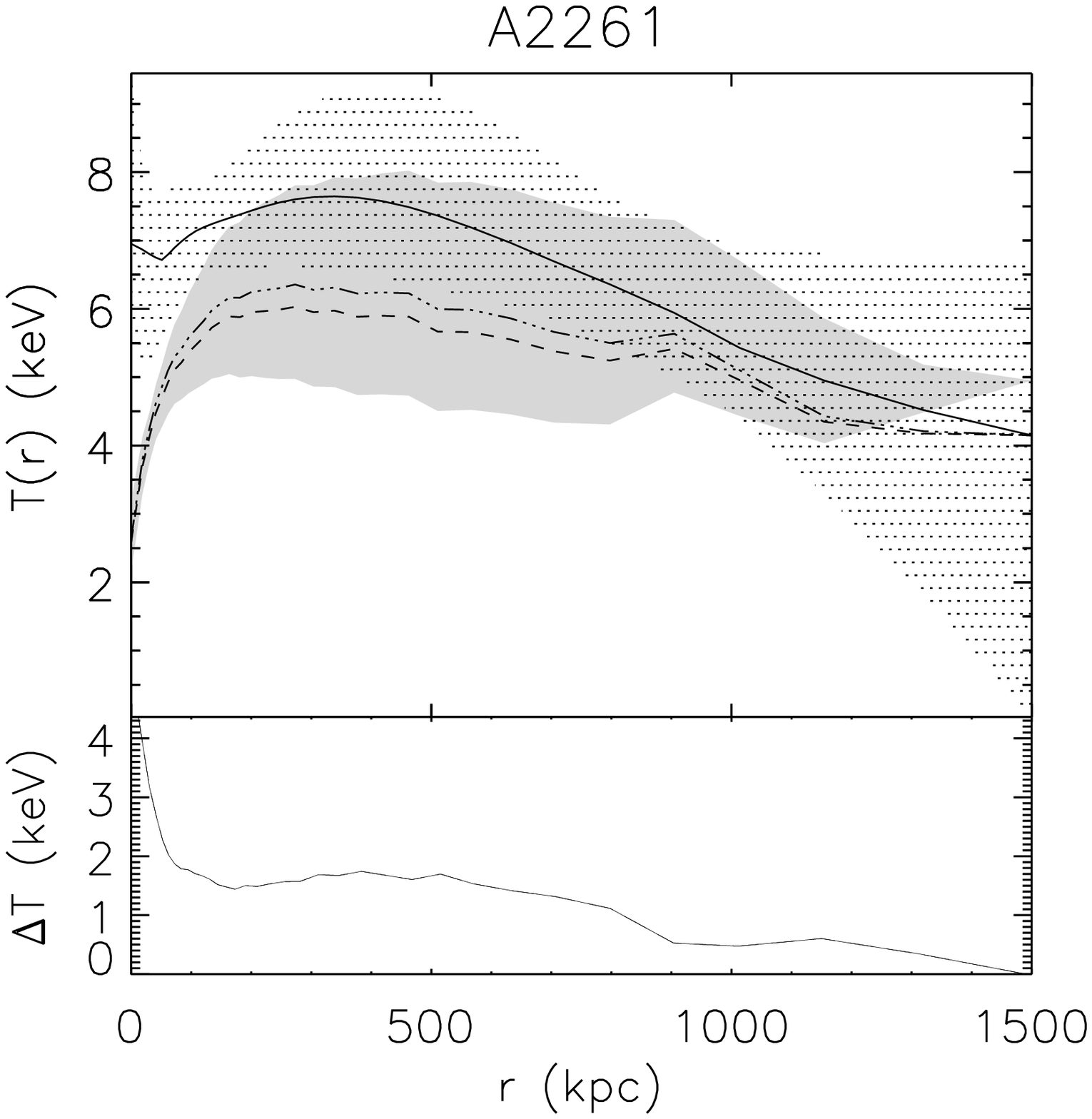 ,width=0.35\textwidth}
\psfig{figure=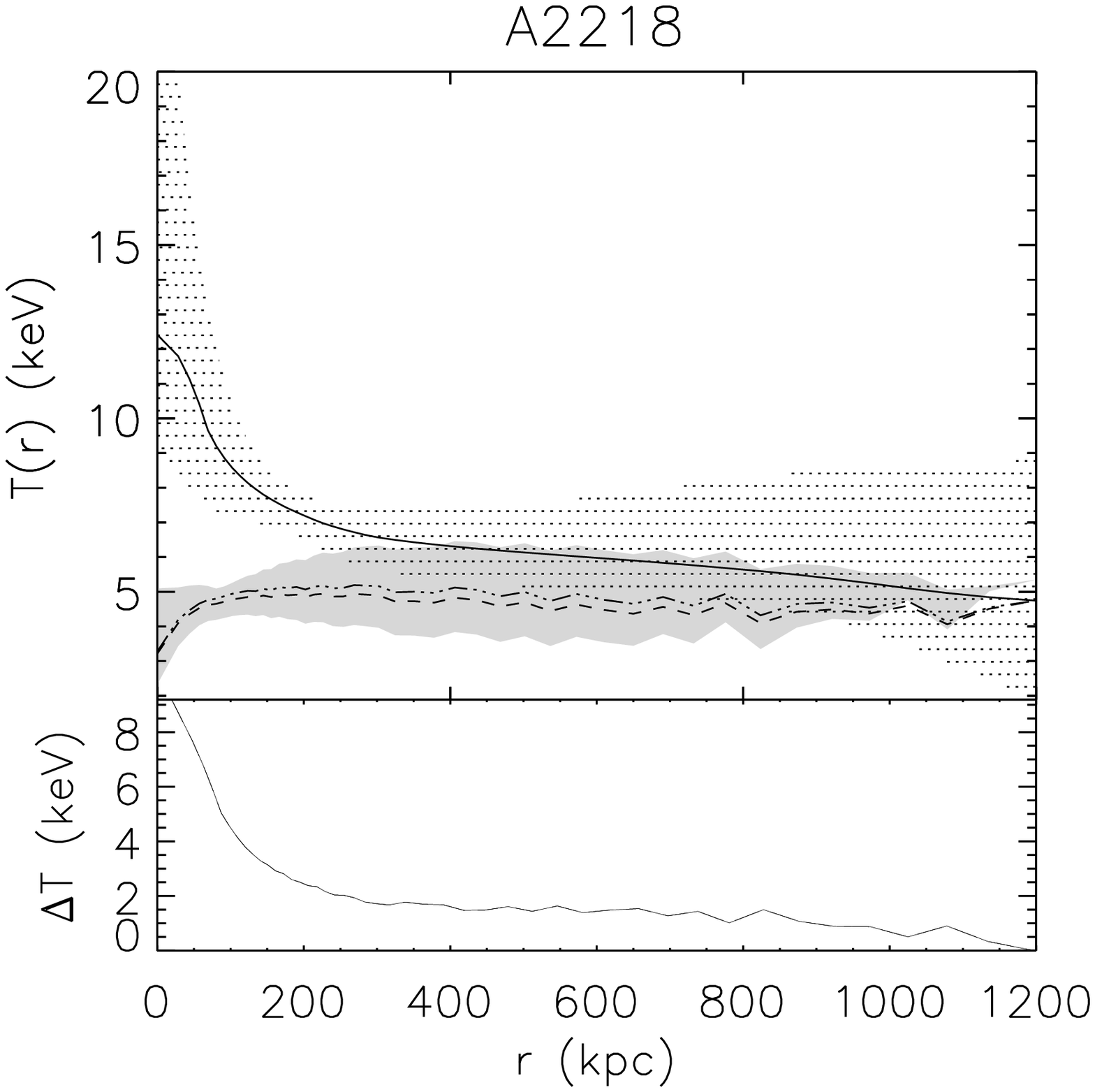,width=0.35\textwidth}
}
\caption[]{Temperature profiles of the gas (solid line) and of the dark matter (dashed and dot-dashed line for $\beta_m=0$ and $\beta_m=0.4$, respectively). The error bands are represented by the gray shaded region for the gas, and hatched region for the DM for the case where $\beta_m=0$. The clusters are A1835 (SCC), A2261 (ICC), and A2218 (NCC), from the left to the right.}
\label{entps3b}
\end{figure*}

We solve eq.~\ref{eq:Jeans1} for $\beta_m=\{0,0.4\}$, corresponding to the case of isotropy of the 
DM and to the central value of the above-mentioned range, respectively, to recover $\sigma_{r}^2$ 
and therefore $T_{\rm DM}$. As boundary condition in eq.~\ref{eq:Jeans1}, we assume $T_{\rm DM}$ equal to $T_{\rm gas}$ at $R_{\rm spec}$. 
We have checked that uncertainties on the DM temperature assumption at $R_{\rm spec}$ are almost negligible 
on the DM temperature profile in the inner and central regions ($R\lesssim R_{2500}\sim 0.25 R_{200}$),
being $R_{\rm spec}\sim 0.3-0.5\, R_{200}$, making our results up to $R_{2500}$ reliable and not
affected from the assumed value at the boundary.
The errors are estimated by looking to the regions of the parameter space that satisfy the condition
$\chi^2 - \chi^2_{\rm min} < 2.3$ after the analysis described in Paper I.
Examples of the gas and DM temperature profiles for SCC, ICC and NCC objects are shown
in Fig.~\ref{entps3b}.

Because only the baryonic component is expected to be prone to non-gravitational energy effects in galaxy clusters,
the difference between $T_{\rm gas}$ and $T_{\rm DM}$, $\Delta kT$, is a powerful tool to trace
the thermal history of the ICM. 
We show in Fig.~\ref{entps3b} how $\Delta kT$ varies as a function of the radius. 
The NCC clusters show a clear trend of $\Delta kT$, with values always greater than zero: 
$\Delta kT \approx 1-2 \,\rm{keV }$ outside the central region ($\gesssim  200-400$ kpc)
and it is a few keV in the inner region. 
Near the cluster observed boundary, DM anisotropies might make $T_{\rm DM}$ roughly in agreement with $T_{\rm gas}$,
even though large statistical errors are present and our boundary condition holds. 
A similar trend is observed in ICC clusters, where a less significant disagreement between $T_{\rm gas}$ and $T_{\rm DM}$
is however observed. In SCC clusters, on the contrary, $T_{\rm gas}$ is well in agreement 
with $T_{\rm DM}$, especially in the inner and central regions.

We notice here that the strong negative evolution measured in the scaling relations between $y_{\Omega}$ and
the X-ray/SZ quantities presented in Paper I, where $y_{\Omega}$ is the integrated Compton parameter over
a fixed angular distance, supports the observed radial behaviour of $\Delta kT$.
Indeed, the measured SZ effect within a fixed angular size samples larger physical region at higher redshifts.
This indicates that the effect of non-gravitational processes is relatively more pronounced 
if the SZ flux is measured within smaller physical radii, indicating the physical scale over which
the non-gravitational processes are more relevant.
When we perform, instead, the same analysis integrating the Compton parameter within a physical radius 
(as done with the quantity $y_{\Delta}$), we observe definitely lower negative evolution.

We have also estimated the global excess of energy $\Delta U_{2500}$ defined in this way:
\begin{equation}
\Delta U_{2500}=\int_0^{R_{2500}} \frac{3}{2} \Delta kT(r)\, n_{\rm gas}(r) \,4\pi r^2 \; d\,r
\label{eq:dm3}
\end{equation}
We find that $\Delta U_{2500}\gesssim 10^{62} \rm{erg}$ in NCC sources (corresponding to about 15-20 per cent of the total thermal energy), that is a factor between 4 and 10 higher than the measured excess in SCC clusters.

We refer to Sect.~\ref{snnen} for a discussion of the observational evidence presented in these two last sections.

\subsection{Relations between gas entropy and metallicity}\label{metallicity}

The ICM iron mass is a key observable to constrain the cumulative past star formation history in galaxy clusters. 
Its relations with other observables such as the cluster optical light, total cluster mass, stellar mass and 
gas entropy, together with its redshift evolution, allow to study the enrichment processes. Moreover, while the production of metals is linked to processes of star formation, its radial profile is determined by different physical processes, such ram-pressure stripping, galactic winds powered by supernovae and AGN activity, merger mechanism \citep{gnedin1998}.

Following the work of \cite{degrandi2004} on local clusters, we present  measured iron abundances in the ICM, their evolution with the redshift $z$ and their correlation with the entropy. 
We adopt the solar abundance ratios from \cite{anders1989} with $Z_\odot={\rm Fe/H}=4.68\times10^{-5}$ by number.

\begin{figure*}
\hbox{
\psfig{figure=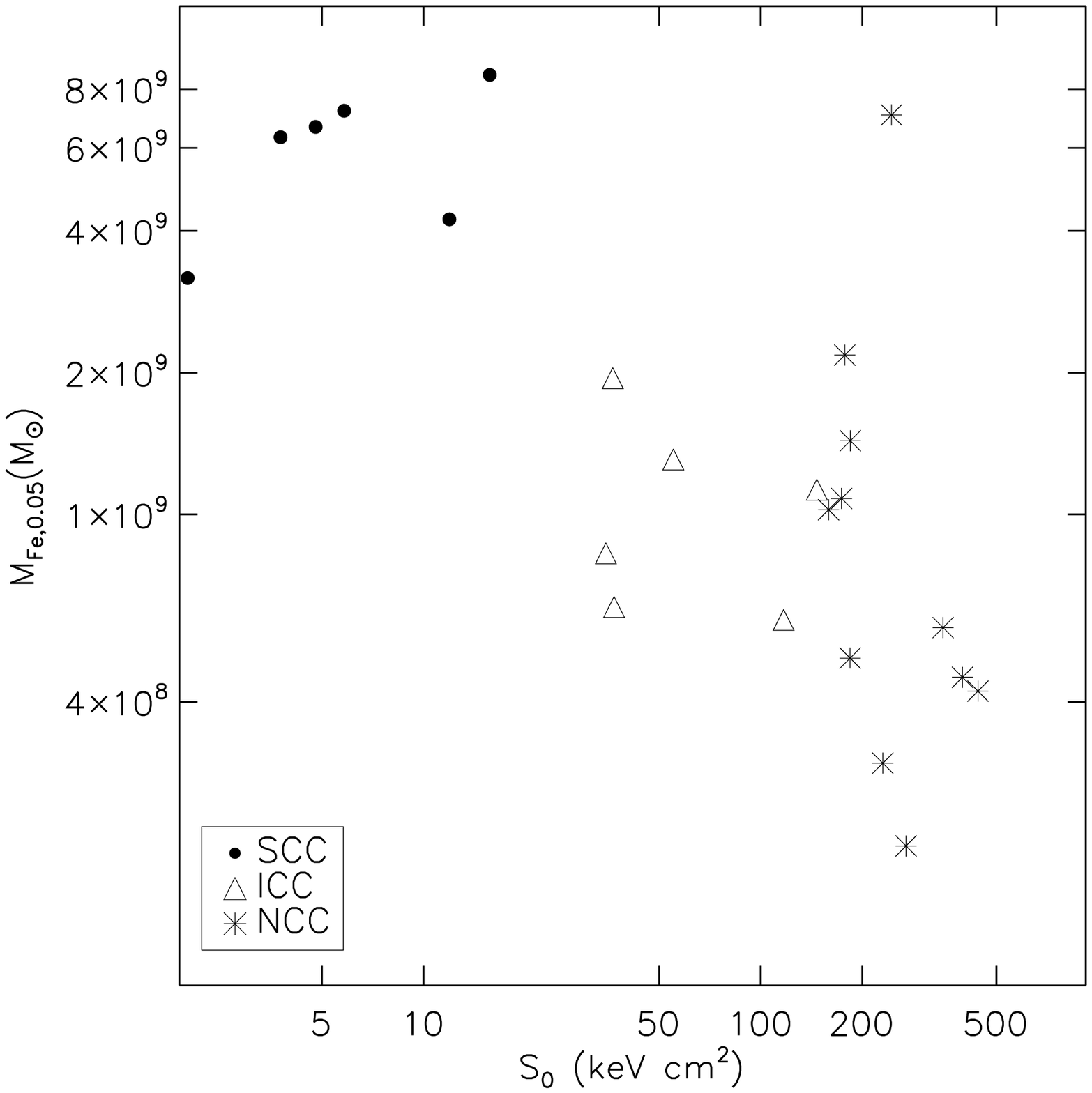,width=0.5\textwidth}
\psfig{figure=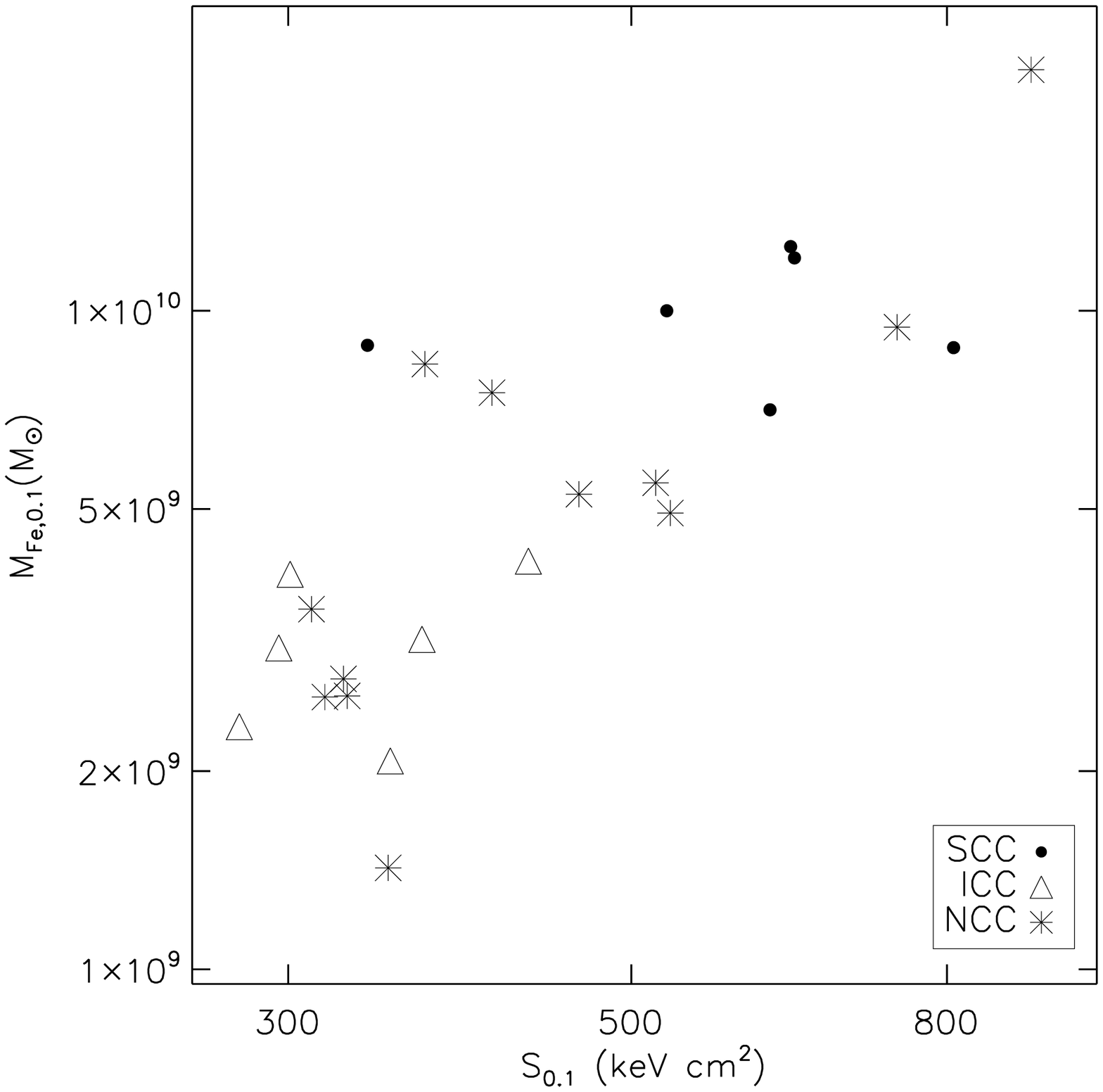,width=0.5\textwidth}
}
\caption[]{
Correlation between $M_{\rm Fe,0.05}$ and the entropy pedestal $S_0$ (left),
$M_{\rm Fe,0.1}$ and $S_{0.1}$ (right).}
\label{entps3bw}
\end{figure*}

We first have determined measures of projected metallicity profiles $Z_{\rm Fe}=Z_{\rm Fe}(r)=n_{\rm Fe}/n_{\rm H}$, (in  units of $Z_\odot$,  that is the solar abundance of iron), where  $n_{\rm Fe}$ and $n_{\rm  H}$ are  the iron and hydrogen densities (by  number) respectively. Notice $Z^{\rm proj}_{\rm Fe}$ has been integrated up 
to $R_{\rm spec}$ without masking the cooling region, to compare our results with the literature available.
In our sample, we find hints of possible negative evolution with redshift, with Spearman's rank coefficient of $r_{\rm s}=-0.12$ for 22 d.o.f (probability of null correlation $p=0.59$), in rough agreement with \cite{balestra2007}, whose sample covers a wider range of $z$.
After the deprojection of the spectral results (see Paper I), we have calculated the iron mass enclosed within a sphere  of radius  $R$ by integrating the iron mass density, $\rho_{\rm Fe}$, over the cluster volume. The total iron mass in solar units can be then written as:
\begin{equation}
M_{\rm Fe} (<R) = 4\pi A_{\rm Fe} m_{\rm H} {Z_\odot
\over M_\odot}~ \int_0^R Z_{\rm Fe}(r) ~n_{\rm H}(r)~ r^2 dr ,
\label{eq:mfe}
\end{equation}
where $A_{\rm Fe}$ is the atomic weight of iron and $m_{\rm H}$ is the atomic unit mass.
To integrate the observed  profiles at any radius, we have linearly interpolated the metallicity mass profiles within overdensities $R_{\delta}=0.05\, R_{200}$, $R_{\delta}=0.1\, R_{200}$ and $\Delta=2500$, which roughly correspond to $100-150$, $200-300$ and $400-600$ kpc for our sample, respectively.

No significant evolution with $z$ of $M_{\rm Fe} (<R_{2500})$ (probability of null correlation $p=0.94$) and $M_{\rm Fe} (<R_{0.05})$ ($p=0.52$) is measured.
Instead, we observe a strong segregation between SSC, ICC and NCC sources, with SCC clusters that tend 
to have higher metallicity mass by a factor of $\sim2$ within $R_{2500}$ and by an order of magnitude
within $R_{0.05}$, which roughly corresponds to the cooling region.
The iron mass excess associated with cool core regions could be entirely produced by the brightest cluster galaxy 
(BCG), which is always found at the centre of cool core clusters, via SN- or AGN- induced winds \citep{degrandi2004}.
Moreover, we confirm the existence of a correlation between $M_{\rm Fe,\delta}$ and $S_{\delta}$. 
In Fig.~\ref{entps3bw}, we present the correlation between the $M_{\rm Fe,0.05}$ as a function of the 
entropy pedestal $S_0$ (see Sect.~\ref{snnen00}): we can see an anti-correlation between 
the two quantities, as expected in a picture where the cooling is the likely predominant physical process in the cooling region. 
Enrichment from recent Supernovae type Ia in the cD galaxies can explain the central metal abundance excess
observed in cooling core clusters \citep{degrandi2004,boehringer2004}.
On the contrary, outside the cooling region, we observe again nearly self-similar relation
between $M_{\rm Fe,0.1}$ and $S_{0.1}$ (see Fig.~\ref{entps3bw}) 
as we have seen in Sect.~\ref{scales1} for the $S_{\delta}-T_{\rm ew}$ relation.

Assuming a synthesized iron mass per SNIa event $m_{\rm Ia}$ of 0.74 $M_{\odot}$ \citep{nomoto1997}
and an energy output of $10^{51}\rm{erg }$, we estimate that $1-3\times 10^{10}$ SNIa events in the region 
inside $R_{2500}$ are required to produce $M_{\rm Fe,2500} \gesssim 0.8\times 10^{10} M_{\odot}$
observed in NCC clusters. This number of SNe corresponds to a global energy output of $1-3\times 10^{61} \rm{erg }$ 
over the entire lifetime of the cluster, that is lower by a factor 2-4 than the excess of energy
$\Delta U_{2500}$ estimated in Sect.~\ref{snnen0} ($\Delta U^{\rm NCC}\sim 10^{62} \rm{erg}$),
suggesting the action of other sources of non-gravitational energy to fully account for this observed excess.

\section{Discussion}\label{snnen}

The main results emerging from our study of the entropy profiles in hot ($kT_{\rm gas} > 6$ keV) galaxy clusters at $z>0.1$ are that these profiles, although similar in the outskirts where they behave as a power-law with slope $1.0-1.2$, are remarkably discrepant in the central regions, with SCC objects that show a power-law behaviour down to the innermost spatially resolved regions and NCC clusters having profiles that flatten to a constant value at $r<0.3 R_{2500}$ (Fig.~\ref{entps1}). Accordingly, the comparison between gas and dark matter temperature profiles (Fig.~\ref{entps3}) reveals that SCC clusters do not present any significant energy excess at any radius, whereas ICC and, more dramatically, NCC objects show $\Delta E=3/2 \Delta kT$ larger than few keV in the cooling region and above. Note that the situation near the boundary of the sources is unclear, because the statistical errors are very large, the effect of possible anisotropies in the DM are there more prominent and we are assuming some constraints on the dark matter temperature at $R_{\rm spec}$.
This excess of energy with respect to the 'gravitational energy floor' associated to the DM 
temperature profile can be interpreted as an indication of the presence of some form of non-gravitational energy that can constrain the mechanisms affecting the ICM thermal history and the observed entropy profiles. 
Indeed, in agreement with $\Delta E\approx 0$ over the entire radial range, we observe that SCC clusters have very low entropy pedestal values $S_{0}$ of few keV cm$^2$, while the higher and more scattered values of $S_{0}$ measured in ICC and, particularly, in NCC systems can be justified by an injection of energy $\Delta E$ of 1-10 keV, that, distributed over scales $\lesssim 100-300$ kpc, explains also their flatter entropy profiles. 
The regular behaviour of the entropy profiles outside $0.1\,R_{200}$ is also in agreement with the fact that $\Delta E$ is low at these radii, 
where we have to consider the limitations of our analysis near $R_{\rm spec}$ as mentioned above. 
This scenario is also supported from our results on the $S-T$ relation, where we observe an higher normalization, 
more significant in the inner regions ($\delta=0.1$; see Table~\ref{entps1a}) of the SCC subsample with respect 
to ICC and NCC sources. We note hints of larger entropy values at higher redshift when our measurements in CC clusters
are compared to the best-fit results obtained in nearby samples, with a more significant deviations observed
at $0.1\,R_{200}$ than at $0.3\,R_{200}$, suggesting that cores in our CC objects are not yet well defined 
from the radiative processes. 
Moreover, the observed mild differences in the slopes of the entropy profile, with $\alpha$ that becomes slightly higher 
by considering NCC, ICC, and SCC sources, respectively ($\alpha^{\rm SCC}=1.18\pm 0.11$, $\alpha^{\rm ICC}=1.07\pm 0.16$ 
and $\alpha^{\rm NCC}=0.95\pm 0.21$ for the powerlaw model) can be explained by looking at the temperature and density profiles (Figure \ref{entps3}), which are a bit flatter for non-cooling core systems: this trend can be justified with small energy excess ($\Delta E \sim$ 1-2 keV) at large scale in the NCC objects compared to the NCC and SCC ones. The radial behaviour of $\Delta E(r)$ is also confirmed by the analysis made in Paper I, where we noticed a strong 
negative evolution in the $y_{\Omega}-$X-ray and $y_{\Omega}-$SZ scaling relations (see Sect. \ref{snnen0}).

All our systems are the products of the hierarchical scenario, how is suggested from the similar behaviour of the 
gas temperature, density, entropy and dark matter profiles in the regions above the cores. 
On the contrary, the cooling region characterizes SCC, ICC, and NCC systems. In particular, continuous interplay between cooling and some form of (pre-)heating can explain the variety of the properties observed, with SCC dominated from the cooling phase and, on the other end, NCC still subjected to
some effects of heating. 

Theoretical models must predict the magnitude of the observed $\Delta E(r)$, and the impact of the non-gravitational processes associated to this excess in the central regions. These models fall into three main classes: preheating, where the gas collapsing into the dark matter potential well is preheated by some mechanism, before clusters were assembled at an early epoch \citep{kaiser1991,balogh1999,tozzi2001,Borgani2005}; local heating by AGN activity, star formation or supernovae \citep{bialek2001,brighenti2006,babul2002,borgani2002}; cooling, which seems to be able to remove low-entropy gas in the centre of the clusters, producing a similar effect to non-gravitational heating \citep{bryan2000,muanwong2002,Borgani2004b}. 
Hereafter, we review the main characteristics of these models and discuss how they are consistent with our observational constraints.

\subsection{Preheating models}\label{preheat}

Models of pre-heating, where a constant energy input is injected either prior of the cluster collapse 
($0.1-0.3$~keV per particle, e.g. \cite{navarro1995,tozzi2001}) or after the cluster formation 
($1-3$~keV per particle (e.g. \cite{metzler1994,loewenstein2000,wu2000,bower2001}, 
could justify, only partially in NCC objects, the observed magnitude of $\Delta E(r)$, but not its radial behaviour. 
Nevertheless, as pointed out by \cite{Borgani2005} by studying hydrodynamical simulated clusters, there is no possibility to inject a large quantity of energy per particle ($\lesssim 1$ keV), unless a large isentropic core is produced in the entropy profile, core that is not observed in our profiles in agreement with other work \citep{ponman2003,pratt2003,pratt2005}. 

\cite{ponman2003} suggest that any raise of the temperature and/or decrease of the density in the gas
inside the primordial structures due to preheating can get largely raised by the accretion shock.
Following the model of \cite{dossantos2002}, \cite{ponman2003} estimated that a mild raise of the entropy of the 
gas confined to filaments ($\sim 10-100 \,\rm{keV \,cm^2}$, corresponding to a temperature of $\sim 10^{-1}$ keV) 
can be boosted by the accretion shock to the observed value of $S$ ($\sim 100-1000 \,\rm{keV \,cm^2}$). 
They point out that an interplay between shock and smoothing of the primordial gas due to a preheating
can justify the observed properties of the gas entropy, given the above upper limits on the energy budget of the preheating
and being the slope of the entropy profile close to the value predicted from shock heating.

Preheating prior of the cluster collapse should be a energetically favorable mechanism compared to {\it in situ} heating to cast further energy into the gas before it is concentrated in the gravitational potential well of the DM halo, since less energy is required to increase the entropy of the gas by a given amount when its density is lower as in the filaments. In fact we observe that in the shock dominated collapse scenario, a mild injection of energy through preheating can greatly amplify the final energy $E_{\rm fin}$ of the post-shocked particles, being $E_{\rm fin} \propto E_{\rm in}$, with $E_{\rm in}$ the initial energy. \cite{Borgani2005} show that smoothing the accretion pattern by preheating in the case of simulations without radiative physics amplifies the entropy generation out to the radius where the accretion shock acts. Nevertheless, the effect seems to be substantially reduced when cooling is also taken into account.

However, our estimates of $\Delta E(r)$ show not a constant profile but instead a declining one outwards.
This behaviour can not be explained by any preheating mechanism, either prior or after cluster collapse,
even though entropy is amplified through subsequent shock heating.
Results for X-ray bright nearby objects by \cite{pratt2006} support this conclusions,
because their scaled entropy profiles show increasing scatter in the inner regions, 
with a dispersion ($\sim 60$ per cent) definitely higher than the value found in simulations 
including filamentary preheating ($\sim$ 30 per cent, \cite{voit2005a}).

\subsection{Heating models}\label{heat}

In principle, heating can amplify the boosting of the entropy out to the radius where accretion shocks are taking place, especially in low mass system, since they are accreted by smaller subhalos where the gas is more smoothed by the extra heating. As pointed out by \cite{Borgani2005}, local heating due to star formation activity is not able alone to prevent overcooling and to reproduce the predicted star formation as low as measured \citep{muanwong2002} and the observed entropy profile: maybe we have to appeal to further sources of non-gravitational energy, like AGN, not taken into account in such hydrodynamical simulations, or different physical mechanisms to distribute the energy inside the ICM. 

The need of this further source of non-gravitational energy is also confirmed by the analysis made in Sect.~\ref{metallicity}, where we observe that the number of supernovae we require to reproduce the observed metallicity is not able to account for all the excess of energy $\Delta U_{2500}$.

A gentle, transonic heating process, such as the weak shocks detected in the Perseus cluster \citep{fabian2003b}, can provide a framework by which one can explain all the observed properties, like the flattening of the entropy profile in the innermost regions ($\lesssim$ a few tens of kpc) even of SCC clusters, and the  spikes of metallicity measured in the centre of SCC sources (see Fig.~\ref{entps3bw}). Weak shocks are indeed likely not able to prevent metals' accumulation in the innermost regions.
\cite{donahue2005,donahue2006} pointed out that the central cooling time of the SCC galaxies ($\sim 10^8$ yr) is consistent with the time scale of the activity of radio sources ($\sim \rm {a \; few}\, 10^7$ yr) at the centre of clusters.
Energy casted by the radio jet ($\sim 10^{45}$ erg/s) can then produce the observed flattening of the entropy
profiles on scale of a few tens of kpc. Gasdynamical models of jets flows proceeding from a central supermassive
black hole and entering surrounding gas may heat the ICM by casting mass and energy outwards till scale
$\sim$ hundreds of kpc, possibly lowering the cooling rate \citep{brighenti2006}.
Nevertheless the above picture does not explain the excess of energy $\Delta E$ in the regions outside the core, 
where we have probably to require some other form of heating, like e.g. shocks induced from merging activity.

Whatever sources of non-gravitational energy we have to appeal, they must fuel energetically the ICM 
in such a way to reproduce the magnitude and the radial behaviour of $\Delta E(r)$,
casting energy (in the order of few keV) and metals preferentially in the innermost regions.

\subsection{Cooling models}

Cooling plays a key role to explain the observed excess of energy $\Delta E(r)$. In fact, $\Delta E(r)$ is higher in the central regions moving from SCC to ICC and to NCC systems.
Cooling can easily account for this trend through radiative losses of the accumulated thermal energy.
In the SCC clusters, $T_{\rm gas}$ is roughly equal to $T_{\rm DM}$ suggesting either that a perfect balance between 
cooling and heating is established, permitting the radiative losses of the only amount of energy in excess
with respect to the one associated to the DM, or that heating is episodic and we are observing structures
in their undisturbed phase.

In a similar manner, \cite{voit2002} have argued that the entropy responsible for the break of the self similarity is not a global property of the ICM, but rather a property set by radiative cooling: 
they point out that the observed entropy value at the core radii of groups and clusters is 
near to the entropy at which $t_{\rm cool} \approx t_{\rm Hubble}$.

A model to explain the observed features in the entropy profile is that proposed by \cite{voit2001}. 
They show how cooling and supernovae heating act to eliminate high compressible gas with $S<S_{\rm c}$, 
being $S_{\rm c}$ the cooling threshold, from the X-ray phase. Much of the condensation and the feedback 
is prior of the epoch of clusters' formation, balancing these processes reciprocally:
more the cooling is effective, more the star formation is active with release of energy to the ICM
and consequent reduction of the cooling itself. This picture is likely not wholly adequate, leading to a 
very large isentropic core in the entropy profiles, which are not observed.

\section{Summary and conclusions}\label{conclusion}

We have presented {\em Chandra} observations of the entropy profiles and scaling properties of a sample of 24 galaxy clusters spanning the redshift range 0.14--0.82 and classified accordingly to their central cooling time in strong (SCC), intermediate (ICC) and non-cooling core (NCC) systems. We have performed a spatially resolved spectral analysis and recovered the gas density, temperature $T$ and entropy $S$ profiles at high spatial resolution and in a non-parametric way. 
We have shown that those entropy profiles are remarkably similar outside the central regions with a typical entropy level at $0.1 \, R_{200}$ of $100-500 \,\rm{keV \,cm^2}$,  
and have a central entropy plateau covering a wide range of values ($\sim$ a few$-200\,\rm{keV \,cm^2}$), 
with the highest values associated to NCC objects.
The CC clusters show larger values of the entropy than the one measured in nearby luminous systems,
with a more significant deviations observed at $0.1\,R_{200}$ than at $0.3\,R_{200}$ with respect to the best-fit
results in Pratt et al. (2006), suggesting that the core in our CC objects are not yet well defined from the cooling processes.

We have studied the radial behaviour of the temperature of the gas ($T_{\rm gas}$) and of the dark matter ($T_{\rm DM}$). We have found that $T_{\rm gas}$ is always higher than $T_{\rm DM}$: for the SCC clusters, the difference of temperature $\Delta kT=kT_{\rm gas}- kT_{\rm DM}$ is negligible, while it is large for the non-cooling core clusters (up to $\sim$10 keV per particle), with $\Delta kT(r)$ that declines outwards.

We conclude that none of the models of (pre-)heating and cooling discussed in literature is able to explain alone 
the observed trends of the entropy profiles and of $\Delta E(r)$. 
Likely, we have to require an interplay of these processes.
A scenario with shock dominated collapse and preheating in the primordial filaments might account for most of the
extra-gravitational energy, as confirmed by the slopes of the entropy profiles near the theoretical value
of 1.1 expected in the accretion shock picture \citep{tozzi2001}. As described in Sect.~\ref{preheat}, 
this should be an energetically favorable mechanism compared to the {\it in situ} heating, 
amplifying significantly the final energy $E_{\rm fin}$ of the post-shocked particles.
On the other side, gentle, sub-sonic heating processes, e.g. supported from AGN's jets, can account for
many of the observed properties, but not for excess of energy still measured outside the core
(see Sect.~\ref{snnen0}).

The relaxed environment of SCC clusters is required to enhance the central metal
abundance and total iron mass in correspondence of low entropy regions (see Sect.~\ref{metallicity}).
There, the cooling is so efficient to remove on short time scale the excess of energy per particle of the ICM,
permitting (i) $T_{\rm gas}$ to approach the dark matter value $T_{\rm DM}$ and
(ii) the other physical parameters used in describing the entropy profile, like, e.g., $S_0$, $r_{\rm break}$, to vary.

Overall, the different observed behaviour of the entropy profiles of SCC, ICC and NCC massive clusters suggest 
that we are observing the end products of the hierarchical model for structure formation. 
They represent different stages of the relative relevance of heating and cooling in regulating the feedback that shapes 
the ICM distribution: galaxy clusters are identified either as NCC objects when heating, probably due to a residual merging 
activity and feedback from AGNs triggered from the merger itself, is predominant, or as SCC systems when the radiative losses 
are energetically prominent, being ICC objects an intermediate case between the two.

\section*{acknowledgements}
We thank the anonymous referee for a careful reading of the manuscripts and suggestions that have improved the presentation of our work. We thank G. W. Pratt, A. Finoguenov, R. Piffaretti and F. Brighenti for useful discussion. We acknowledge the financial support from contract ASI-INAF I/023/05/0 and from the INFN PD51 grant.


\end{document}